\documentclass[11pt,letterpaper]{article}

\usepackage{jheppub}
\usepackage{braket}
\usepackage{amsmath}
\usepackage{amsfonts}
\usepackage{amssymb}
\usepackage[force]{feynmp-auto}
\usepackage{graphicx}
\usepackage{epsfig}
\usepackage{color}
\usepackage{hyperref}
\usepackage{enumitem}
\usepackage{multirow}
\usepackage{todonotes}
\usepackage[normalem]{ulem}

\def\ba{\begin{eqnarray}}
\def\ea{\end{eqnarray}}
\def\be{\begin{equation}}
\def\ee{\end{equation}}
\def\bea{\begin{eqnarray}}
\def\eea{\end{eqnarray}}
\def\bef{\begin{flalign}}
\def\eef{\end{flalign}}
\def\nn{\nonumber}
\def\d{\mathrm{d}}
\def\vk{\vec{k}}
\def\vp{\vec{p}}
\def\vx{\vec{x}}

\def\vq{\vec{q}}

\def\hPhi{\hat{\Phi}}
\def\la{\langle}
\def\ra{\rangle}
\def\({\left(}
\def\){\right)}
\def\[{\left[}
\def\]{\right]}
\def\<{\left\langle}
\def\>{\right\rangle}

\def\tt{\textcolor{violet}}

\title{\center{Loop corrections in Minkowski spacetime away from equilibrium. Part I. Late-time resummations}}

\author[a]{Spasen Chaykov,}
\author[a]{Nishant Agarwal,}
\author[b]{Sina Bahrami,}
\author[c]{and R. Holman}

\affiliation[a]{Department of Physics and Applied Physics, University of Massachusetts, Lowell, MA 01854, USA}
\affiliation[b]{Institute for Gravitation and the Cosmos, The Pennsylvania State University, University Park, PA 16802, USA}
\affiliation[c]{Minerva University,
     14 Mint Plaza, Suite 300, San Francisco, CA 94103, USA}

\emailAdd{spasen\_chaykov@student.uml.edu}
\emailAdd{nishant\_agarwal@uml.edu}
\emailAdd{sb933@cornell.edu}
\emailAdd{rh4a@andrew.cmu.edu}

\abstract{Loop corrections to unequal-time correlation functions in Minkowski spacetime exhibit secular growth due to a breakdown of time-dependent perturbation theory. This is analogous to secular growth in equal-time correlators on time-dependent backgrounds, except that in Minkowski the divergences must not signal a real IR issue. In this paper, we calculate the late-time limit of the two-point correlator for different massless self-interacting scalar quantum field theories on a Minkowski background. We first use a late-time version of the in-in path integral starting in the vacuum of the free theory; in this limit, the calculation, including UV renormalization, reduces to that in in-out. We find linear or logarithmic growth in time, depending on whether the interaction strength is dimension-one or dimensionless, respectively. We next develop the Weisskopf-Wigner resummation method, that proceeds by demanding unitarity within a truncated Hilbert space, to calculate the resummed correlator and find that it gives an exact exponentiation of the late-time perturbative result. The resummed (unequal-time) correlator thus decays with an exponential or polynomial time-dependence, which is suggestive of `universal' behavior that depends on the dimensions of the interaction strength.}

\keywords{Non-equilibrium field theory, Renormalization and regularization, Effective field theories}
\arxivnumber{2206.11288}

\begin{document}

\maketitle


\section{Introduction}
\label{sec:intro}

The techniques of out-of-equilibrium quantum field theory have successfully been applied to the calculation of observables in an inflationary Universe. Loop corrections to correlation functions in inflation, and even in the simpler case of a massless self-interacting scalar field on the Poincar\'{e} patch of de Sitter spacetime, however, are known to exhibit secular growth and IR divergences; see \cite{Seery:2010kh,Tanaka:2013caa,Hu:2018nxy} for detailed reviews. It has been argued in \cite{Burgess:2009bs,Boyanovsky:2012qs,Chen:2016nrs} that the divergences may, in some cases, lead to a dynamical generation of mass for the field. On the other hand, other works have argued that the infrared pathology in de Sitter can be cured within a stochastic or Fokker-Planck-like approach \cite{Gorbenko:2019rza,Baumgart:2019clc} and the curvature perturbation in inflation remains massless due to a subtle cancellation between various one-loop diagrams \cite{Pimentel:2012tw}.

While part of the complication is unique to inflation/de Sitter, where one is calculating time-dependent correlators on a time-dependent background, some of the divergences must arise simply from the use of time-dependent perturbation theory. In this paper, we thus take a step back and work in Minkowski spacetime, so that the background is time-independent and the physical wavelength of modes is not a function of time either. We consider different interacting theories of a massless scalar field $\phi(\vx,t)$ in Minkowski, namely a $\lambda\phi^3$ interaction in 4D and 6D and a $\lambda\phi^4$ interaction in 4D. We take particular care in renormalizing the theory, choosing counterterms that respect the background Lorentz symmetry. We assume that the interaction is switched on adiabatically in the infinite past, in which case the correlators that we are interested in can be obtained using the standard techniques of (equilibrium) in-out perturbation theory found in most textbooks on quantum field theory. The finite-time calculation, on the other hand, can be done using (out-of-equilibrium) in-in perturbation theory \cite{Schwinger:1960qe,Mahanthappa:1962ex,Bakshi:1962dv,Kadanoff:1962,Bakshi:1963bn,Keldysh:1964ud,Jordan:1986ug,Calzetta:1986ey}, and is carried out in a companion paper \cite{Chaykov:2022pwd}.

As will become clear in the paper, loop corrections to equal-time correlations in Minkow- ski, calculated in the vacuum of the free theory, do not exhibit secular growth. Those to unequal-time correlations, however, {\it do} grow with time. At second order in perturbation theory, we find that the two-point correlator grows linearly in the difference of the two times for $\lambda\phi^3$ in 4D (where $\lambda$ has dimensions of mass) and logarithmically for both $\lambda\phi^3$ in 6D and $\lambda\phi^4$ in 4D (where $\lambda$ is dimensionless). These late-time divergences, however, must be artificial and must disappear in a non-perturbative calculation since field correlations would otherwise grow indefinitely without any time-dependence in the Hamiltonian. As will also become clear in the paper, we should in fact expect the correlator to {\it decay} at late times, due to the decay of a one-particle state.

We next use the Weisskopf-Wigner (WW) method to perform a late-time perturbative resummation of the correlator. The WW method was originally developed to calculate the decay width of an atomic system in quantum electrodynamics \cite{Weisskopf1930}. It was generalized to study the perturbative stability of quantum fields on a de Sitter background in \cite{Boyanovsky:2011xn}, where it was also shown to connect to the dynamical renormalization group approach. The WW method resums various transition amplitudes by demanding unitarity within a truncated Hilbert space. For the correlator calculation that we are interested in, it turns out that we need to perform the resummation twice and consider only those states that are connected to the starting state at first order in the interaction. We find that the resummed result for the unequal-time two-point correlator is an exact exponentiation of the late-time perturbative result and decays exponentially for $\lambda\phi^3$ in 4D and polynomially for both $\lambda\phi^3$ in 6D and $\lambda\phi^4$ in 4D.

The paper is organized as follows. We set up our notation and the problem in section\ \ref{sec:setup}. In section\ \ref{sec:inout}, we explain why in-out perturbation theory suffices for the calculation that we are interested in and use it to calculate the unequal-time two-point correlator for different interactions, including the requisite renormalization counterterms. The loop integrals encountered in this section can be found in standard textbooks, but are shown in two appendices to make the paper self-sufficient. In section\ \ref{sec:ww}, we briefly review the WW resummation method and use it to calculate the resummed late-time two-point correlator for different interactions, including the renormalization counterterms. Details of the vacuum state amplitudes encountered in this section are also relegated to two appendices. In section\ \ref{sec:GRWW}, we briefly discuss how the geometric series resummation that can be performed for in-out correlators compares to results obtained using the WW resummation method, and highlight what type of diagrams are resummed by the WW method. We end with a discussion in section\ \ref{sec:disc}.


\section{Setup}
\label{sec:setup}

Consider the Schr\"{o}dinger picture field operator $\hPhi_S(\vx)$ with eigenstates $| \phi(\cdot) \rangle$ and eigenvalues $\phi(\vx)$, so that $\hPhi_S(\vx) | \phi(\cdot) \rangle = \phi(\vx) | \phi(\cdot) \rangle$. The dot here indicates all field configurations and the completeness relation is given by $\int {\cal D} \phi(\vx)| \phi(\vx) \rangle \langle \phi(\vx) | = \hat{1}$. In this paper, we are interested in calculating the connected correlation function ${\rm Tr} \big[ \hat{\rho}(t_0) \hPhi(\vx,t) \hPhi(\vx',$ $t') \big]_c$, or rather its spatial Fourier transform, with $t > t'$, for different self-interacting field theories; here $\hat{\rho}(t_0)$ is the initial density operator for the field, which we choose to coincide with the free theory's ground state, and $\hPhi(\vx,t)$ is the Heisenberg picture field operator. The standard method to do this calculation uses the in-in path integral approach, that we discuss in detail in a companion paper \cite{Chaykov:2022pwd}. In this paper, we restrict to the late-time limit $t_0 \rightarrow -\infty$ or, equivalently, $t,  t' \gg t_0$ and study secular growth in the limit that $t \gg t'$. In the late-time limit, and with our choice of initial state, it turns out that the scattering or in-out approach suffices, as we discuss in the next section.

We specifically consider the action $S[\phi] = \int \d^dx \, {\cal L}[\phi]$ in $d$-dimensional Minkowski spacetime with the following two Lagrangian densities,
\bea
    {\cal L}_3[\phi] & = & -\frac{1}{2} (\partial_{\alpha}\phi)^2 - \frac{1}{2} m^2 \phi^2 - \frac{1}{3!} \lambda \phi^3 - \frac{1}{2} \delta_r (\partial_{\alpha}\phi)^2 - \frac{1}{2} \delta_m \phi^2 - \frac{1}{3!} \delta_\lambda \phi^3 + Y\phi \quad
\label{eq:lag3}
\eea
and
\bea
    {\cal L}_4[\phi] & = & -\frac{1}{2} (\partial_{\alpha}\phi)^2 - \frac{1}{2} m^2 \phi^2 - \frac{1}{4!} \lambda \phi^4 - \frac{1}{2} \delta_r (\partial_{\alpha}\phi)^2 - \frac{1}{2} \delta_m \phi^2 - \frac{1}{4!} \delta_\lambda \phi^4 \, ,
\label{eq:lag4}
\eea
where $(\partial_{\alpha}\phi)^2 = -\dot{\phi}^2 + (\partial_i\phi)^2$ with the dot now indicating a derivative with time, $m$ is the mass parameter, $\lambda$ is a coupling constant, and the terms with $\delta_r$, $\delta_m$, $\delta_{\lambda}$, and $Y$ are the usual counterterms `added' to cancel any UV divergences. We treat the $\lambda$ terms perturbatively as usual, expanding around $\lambda = 0$, and restrict our calculations to second order in perturbation theory, or ${\cal O}(\lambda^2)$. Further, we restrict our calculations to the massless ($m = 0$) case for technical reasons. At this order and with $m$ set to zero, specific counterterms may or may not contribute. Namely, for the one- and two-point correlation function calculations that we are interested in, we do not need to renormalize the vertex and can, therefore, set $\delta_{\lambda}$ to zero ($\delta_{\lambda}$ is zero at second order for the $\lambda\phi^3$ interaction and its contribution at second order to the two-point correlator vanishes in the massless case for the $\lambda\phi^4$ interaction).

We also use the modified minimal subtraction ($\overline{\text{MS}}$) scheme for renormalization and so setting the mass parameter $m$ to zero does not necessarily correspond to having zero physical mass. We are primarily interested in a late-time resummation of the perturbative result, however, and, therefore, do not consider the running of the physical mass or coupling and instead leave our results in terms of an arbitrary renormalization scale $\mu$. We also note that in the case of $\lambda\phi^3$ in 4D, the coupling has dimensions of mass and, especially in the massless limit, the statement that `$\lambda$ is small' is ambiguous. In fact, time-dependent perturbation theory will eventually break down for any $\lambda$, as we find later in the paper. We nevertheless treat the problem perturbatively and infer when such a treatment is valid based on our results.


\section{Calculating late-time correlations using in-out}
\label{sec:inout}

On pushing the initial time $t_0$ to $-\infty$ and choosing $\hat{\rho}(t_0)$ to be $|0\rangle \langle 0|$, $|0\rangle$ being the vacuum of the free theory, the time-ordered correlation ${\rm Tr} \big[ \hat{\rho}(t_0) T \big\{ \hPhi(\vx,t) \hPhi(\vx',t') \big\} \big]$ reduces to the Feynman Green's function in in-out perturbation theory. To see why this works, let us first remove the time-ordering and write the resulting correlation in terms of interaction picture fields as $\big\la \hat{U}_{i,I}^{\dagger}(t,t_0) \hPhi_I(\vx,t) \hat{U}_{i,I}(t,t') \hPhi_I(\vx',t') \hat{U}_{i,I}(t',t_0) \big\ra$, where we use angular brackets to denote the expectation value in $|0\rangle$, $\hat{U}$ is the unitary time evolution operator, the subscript `$i$' denotes the interaction, and `$I$' indicates interaction picture. Let us now choose a new reference time, that we will set to zero, to transform between different pictures. Then, after manipulating the first and last time evolution operators, the correlation becomes $\big\la \hat{U}_{i,I}^{\dagger}(0,t_0) \hPhi(\vx,t) \hPhi(\vx',t') \hat{U}_{i,I}(0,t_0) \big\ra$. Assuming that $\hat{H}_0 |0\rangle = 0$ for the free Hamiltonian, the remaining two time evolution operators can be replaced with full $\hat{U}$'s rather than their interaction picture counterparts. Now setting $t_0$ to $-\infty$ and assuming that the interaction is switched on adiabatically in the past, we can relate $|0\rangle$ to the ground state of the interacting theory, denoted $|\Omega\rangle$, via $|\Omega\rangle = \hat{U}(0,-\infty) |0\rangle$ \cite{Gell-Mann:1951ooy}. The correlation function then reduces to $\langle \Omega| \hPhi(\vx,t) \hPhi(\vx',t') |\Omega \rangle$. Lastly, we can repeat this calculation upon interchanging the fields and obtain the result with time-ordering; the time-ordered correlation is, therefore, identical to that in in-out calculations \cite{Kamenev:2011}.\footnote{In order to make the analogy with in-out perturbation theory complete, we also need to assume that the interaction is switched off adiabatically in the future, so that $\langle 0| \hat{U}(\infty,-\infty) |0\rangle$ is simply a phase factor.}

In the limit that $t_0$ is $-\infty$ and $\hat{\rho}(t_0)$ is $|0\rangle \langle 0|$, we can, therefore, use the standard techniques of in-out perturbation theory. In particular, time-ordered correlation functions can be obtained by taking functional derivatives of the in-out generating functional,
\bea
    Z[J] & = & _{\rm out}\la 0 | 0 \ra_{{\rm in}, J} \nn \\
    & = & \lim_{T \rightarrow \infty(1-i\epsilon)} \int {\cal D} \phi(x) \, {\rm exp} \[ i\int_{-T}^T \d t \int \d^{d-1}x \( {\cal L}[\phi] + J\phi \) \],
\eea
where $x$ is the $d$-dimensional coordinate and $\epsilon > 0$ is infinitesimal, with respect to the external source $J(x)$ that is set to zero at the end of the calculation; see, for example, \cite{Srednicki:2012,Peskin:2015}. We can proceed by making use of perturbation theory as usual, writing the full action as $S[\phi] = S_0[\phi] + S_i[\phi]$, $S_0$ being the free part and $S_i$ the interaction part. Also going to Fourier space,\footnote{We use the Fourier convention $f(\vx,t) \, = \, \int_{\vk} e^{i\vk \cdot \vx} f \big( \vk, t \big) \, = \, \int_k e^{i k \cdot x} f(k)$ with the shorthand $\int_{\vk} \equiv \int \frac{\d^{d-1} k}{(2\pi)^{d-1}}$ and $\int_k \equiv \int \frac{\d^d k}{(2\pi)^d}$ throughout this paper. The $\delta$-function is, therefore, given by $\int \d^{d-1} x \, e^{-i\vk \cdot \vx} = (2\pi)^{d-1} \delta^{d-1} \big( \vk \big)$ and $\int \d^d x \, e^{-i k \cdot x} = (2\pi)^d \delta^d (k)$. We further relegate spatial Fourier indices to subscripts from now on for the ease of notation.} the generating functional can then be written as
\bea
     Z[J] & = & Z_0[0] \, {\rm exp} \left\{ i S_i \[ -i \frac{\delta}{\delta J} \] \right\} {\rm exp} \[ -\frac{1}{2} \int_k J(k) \, G_F(k) J(-k) \] ,
\label{eq:genfunc}
\eea
where $Z_0[0]$ is the free theory generating functional in the absence of external sources. The functional derivatives in $S_i$ generate loop corrections and $G_F(k)$ is the Feynman Green's function of the free theory,
\bea
    G_F(k) & = & \frac{i}{-k^2 -m^2 + i\epsilon} \, .
\label{eq:GF}
\eea
The Fourier transform of the time-ordered correlation function $\big\la T \hPhi(x) \hPhi(x') \big\ra$ is now given by
\bea
    \int \d^d x \int \d^d x' e^{-i (k \cdot x + k' \cdot x')} \big\la T \hPhi(x) \hPhi(x') \big\ra & = & \frac{1}{Z[J]} \( -i \frac{\delta}{\delta J(k)} \) \( -i \frac{\delta}{\delta J(k')} \) Z[J] \bigg|_{J = \, 0} \, . \quad \ \ 
\label{eq:2ptdef}
\eea
The $Z[0]$ in the denominator cancels, as usual, any vacuum diagrams (disconnected diagrams without external sources). We also ensure that the one-point correlation vanishes, so that eq.\ (\ref{eq:2ptdef}) matches with the connected two-point correlation and equals $(2\pi)^d \delta^d (k+k') G_{F,{\rm full}}(k)$. Note that the free theory result is simply $(2\pi)^d \delta^d (k+k') G_F(k)$ and we can thus refer to loop corrections in an interacting theory as corrections to $G_F(k)$. We are finally interested in this paper in the correlation function $\la \hPhi_{\vk}(t) \hPhi_{\vk'}(t') \ra_c$ for $t > t'$, which can be found by taking the inverse Fourier transform of $G_{F,{\rm full}}(k)$ with respect to $k_0$,
\bea
    \big\la T \hPhi_{\vk}(t) \hPhi_{\vk'}(t') \big\ra_c & = & (2\pi)^{d-1} \delta^{d-1} \big( \vk+\vk' \big) \int \frac{\d k_0}{2\pi} \, e^{-ik_0 (t-t')} G_{F,{\rm full}}(k) \, ,
\label{eq:ftfinal}
\eea
and setting $t > t'$ at the end of the calculation. We calculate this at second order in perturbation theory for the Lagrangian densities in eqs.\ (\ref{eq:lag3}) and (\ref{eq:lag4}) in the two subsections below.

\subsection{\texorpdfstring{$\lambda\phi^3$}{lp3} in 4D and 6D}

\begin{figure*}[!t]
\begin{center}
	\includegraphics[scale=0.6]{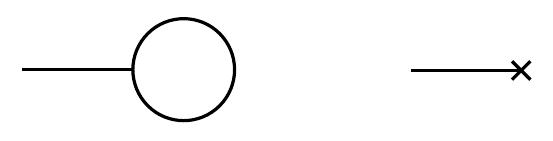}
	\caption{(Left) One-loop contribution to the one-point function in a $\lambda\phi^3$ theory. (Right) The corresponding $Y$ counterterm diagram.}
\label{fig:tp}
\end{center}
\end{figure*}

We first consider a $\lambda\phi^3$ interaction in $d$ dimensions, setting $d = 4$ for the case that $\lambda$ has dimensions of mass or $d = 6$ where $\lambda$ is dimensionless later in the calculation. The first quantity that we need to calculate is the one-point function, which we only need at one-loop order, or at ${\cal O}(\lambda)$, since any other contributions are higher than second order in $\lambda$. We choose the linear counterterm $Y$ such that the one-point function vanishes, which cancels any tadpole contributions to the two-point function as well. The interaction and counterterm diagrams that contribute are shown in fig.\ \ref{fig:tp} and give
\bea
    \int \d^d x \, e^{-i k \cdot x} \big\la \hPhi(x) \big\ra & = & \frac{1}{Z[J]} \( -i \frac{\delta}{\delta J(k)} \) Z[J] \bigg|_{J = \, 0} \nn \\
    & = & - \frac{\lambda}{2} \int_p G_F(k) G_F(p) + Y G_F(k) \, ,
\label{eq:tp}
\eea
which can be made to vanish by choosing
\bea
    Y & = & \frac{\lambda}{2} \int_p G_F(p) \, .
\label{eq:linct}
\eea
The explicit dimensional regularization-based calculation of $Y$ can be found in standard textbooks, but is shown in appendix\ \ref{app:31l} where we find, as usual, that it vanishes in the massless limit. Also note that adding the $Y$ counterterm is equivalent to normal ordering the interaction, $\int \d^d x \, \frac{1}{3!} \lambda \hat{\Phi}^3 \rightarrow \int \d^d x \, \frac{1}{3!} \lambda {\,:}\hat{\Phi}^3{:\,}$. This will be useful in the WW calculation that we perform in section\ \ref{sec:ww}.

\begin{figure*}[!t]
\begin{center}
	\includegraphics[scale=0.6]{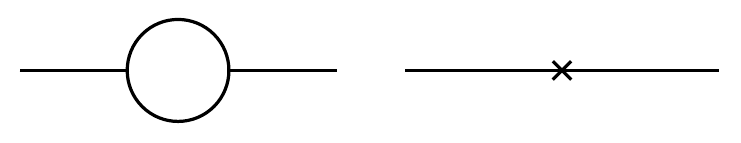}
	\caption{(Left) Connected one-loop contribution to the two-point correlation function in a $\lambda\phi^3$ theory. (Right) The corresponding $\delta_m$ and/or $\delta_r$ counterterm diagrams.}
\label{fig:2p}
\end{center}
\end{figure*}

Having fixed the linear counterterm, we now calculate the one-loop correction to the two-point correlation, which contributes at ${\cal O}(\lambda^2)$. We can obtain this from the generating functional as in eq.\ (\ref{eq:2ptdef}), which leads to the interaction and $\delta_m$ and $\delta_r$ counterterm diagrams shown in fig.\ \ref{fig:2p}. Their contribution to the Feynman Green's function is given by
\bea
    G_{F,1-{\rm loop}}(k) & = & -\frac{\lambda^2}{2} \int_{p} G_F(k) G_F(p) G_F(k-p) G_F(k) - i \( \delta_m + k^2 \delta_r \) G_F(k) G_F(k) \, . \quad
\label{eq:phi32p}
\eea
We only need to find the loop integral
\bea
    I_{\lambda\phi^3}(k) & = & \frac{\lambda^2}{2} \int_p G_F(p) G_F(k-p) \, ,
\label{eq:phi3loop}
\eea
which is also a standard calculation, but is shown using dimensional regularization in appendix\ \ref{app:31l} for both cases of interest, $d = 4$ and $d = 6$. We consider each of these cases in turn below.

\subsubsection*{(i) $\lambda\phi^3$ in 4D}

We first set $d = 4 + \epsilon$, where $\epsilon$ is a dimensional regulator. Inserting the result from eq.\ (\ref{eq:phi3loop4d6d}) into eq.\ (\ref{eq:phi32p}) gives
\bea
    G_{F,1-{\rm loop}}(k) & = & G_F^2(k) \[	-\frac{i\lambda^2}{16\pi^2 \epsilon} + \frac{i\lambda^2}{32\pi^2} \left\{ 2 - \gamma + \ln \( \frac{4\pi\mu^2}{k^2} \) \right\} - i \( \delta_m + k^2 \delta_r \) \] , \quad
\label{eq:2p4dbf}
\eea
where $\mu$ is a parameter with dimensions of mass and can be thought of as a renormalization scale. The $\mu$-dependence would typically be absorbed into renormalized couplings, but we will leave it as is since we are only interested in the time-dependence in this paper. The UV divergence as $\epsilon \rightarrow 0$ can be canceled with the following choice of counterterms,
\bea
    \delta_m \ = \ -\frac{\lambda^2}{32 \pi^2} \[ \frac{2}{\epsilon} + \gamma - \ln (4\pi) \] , \quad \delta_r \ = \ 0 \, ,
\label{eq:d4ct}
\eea
where we have used the $\overline{\rm MS}$ renormalization scheme. We now plug the resulting expression for $G_{F,1-{\rm loop}}(k)$ into eq.\ (\ref{eq:ftfinal}) where we perform the inverse Fourier transform over $k_0$ and choose $t>t'$ without loss of generality.\footnote{We note that in order to perform the inverse Fourier transform, we need to reintroduce the shift $-k^2\rightarrow-k^2+i\epsilon$ in the logarithm in eq. (\ref{eq:2p4dbf}).} We also take the late-time limit such that $t-t'$ is bigger than any timescale in the problem, $\lambda (t-t') \gg 1$, $k (t-t') \gg 1$, and $\mu (t-t') \gg 1$. This gives us a final expression for the connected two-point correlation at one-loop and in the late-time limit,
\bea
    \big\la \hPhi_{\vk}(t) \hPhi_{\vk'}(t') \big\ra_c & = & (2\pi)^3 \delta^3 \big( \vk+\vk' \big) \frac{e^{-i |\vk| (t-t')}}{2|\vk|} \Bigg[ 1 - \frac{\lambda^2(t-t')}{64\pi^2|\vk|} \Bigg\{ \frac{\pi}{2} \nn \\
    & & \qquad - \ i \ln \( \frac{\mu^2(t-t')}{2|\vk|} \) - i - i\gamma \Bigg\} \Bigg] .
\label{eq:2pft4d}
\eea
We see that the result diverges as $\lambda^2 (t-t')/|\vk| \rightarrow \infty$ or $\mu^2 (t-t')/|\vk| \rightarrow \infty$, signaling a breakdown of perturbation theory. We will find in section\ \ref{sec:ww} that the WW resummation method exponentiates the secular term here leading to an exponential decay instead, which is understood as arising from the decay of one-particle states in Minkowski.

We point out that, in principle, one could write a geometric series in all 1PI contributions to the propagator using eq. (\ref{eq:2p4dbf}) at ${\cal O}(\lambda^2)$, find the resummed in-out propagator, and then perform an inverse Fourier transform. This result should also not show a late-time divergence. The inverse Fourier transform, however, is not always doable, for example, the $\ln k^2$ that will result in the denominator in this particular case makes the inverse Fourier transform hard. We can alternatively try a resummation before the $\epsilon$ expansion, essentially using eq.\ (\ref{eq:phi3loopbeforeexp}) for the loop correction, but we were still unable to perform the inverse Fourier transform of the resulting expression. That this should work, however, can be checked through a simple calculation that we discuss in section\ \ref{sec:GRWW}. Our goal here is not to use in-out to its fullest extent, but rather to demonstrate why late-time divergences appear and how to potentially handle them, especially in out-of-equilibrium calculations where a geometric resummation is not feasible. 

\subsubsection*{(ii) $\lambda\phi^3$ in 6D}

We next set $d = 6+\epsilon$, in which case inserting the result from eq.\ (\ref{eq:phi3loop4d6d}) into eq.\ (\ref{eq:phi32p}) gives
\bea
    G_{F,1-{\rm loop}}(k) & = & G_F^2(k) \[ \frac{i k^2 \lambda^2}{384 \pi^3 \epsilon} - \frac{i\lambda^2 k^2}{768\pi^3} \left\{ \frac{8}{3}-\gamma + \log \( \frac{4\pi\mu^2}{k^2} \) \right\} - i\( \delta_m + k^2 \delta_r \) \] . \qquad
\label{eq:2p6dbf}
\eea
The UV divergence can now be canceled with the following choice of counterterms,
\bea
    \delta_m \ = \ 0 \, , \quad \delta_r \ = \ \frac{\lambda ^2}{768 \pi ^3} \[ \frac{2}{\epsilon} + \gamma - \ln(4\pi) \] ,
\label{eq:d6ct}
\eea
where we have again used the $\overline{\rm MS}$ renormalization scheme. Now plugging in the resulting expression for $G_{F,1-{\rm loop}}(k)$ into eq.\ (\ref{eq:ftfinal}) and taking the late-time limit, $k (t-t') \gg 1$ and $\mu (t-t') \gg 1$, yields
\bea
    \big\la \hPhi_{\vk}(t) \hPhi_{\vk'}(t') \big\ra_c & = & (2\pi)^5 \delta^5 \big( \vk+\vk' \big) \frac{e^{-i |\vk| (t-t')}}{2|\vk|} \[ 1 - \frac{\lambda^2}{768\pi^3} \ln\left\{ \mu(t-t') \right\} \] ,
\label{eq:2pft6d}
\eea
which again diverges as $\mu (t-t') \rightarrow \infty$, signaling a breakdown of perturbation theory. We will find in section\ \ref{sec:ww} that the WW resummation method exponentiates the secular term here as well, leading to a polynomial decay instead.


\subsection{\texorpdfstring{$\lambda\phi^4$}{lp4} in 4D}

\begin{figure}[!t]
\begin{center}
	\includegraphics[scale=0.6]{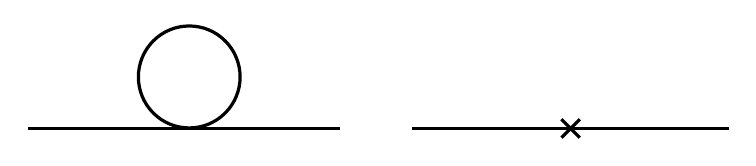}
	\caption{(Left) One-loop contribution to the two-point correlation function in a $\lambda\phi^4$ theory. (Right) The corresponding $\delta_m$ and/or $\delta_r$ counterterm diagrams.}
\label{fig:2p1l}
\end{center}
\end{figure}

We next consider a $\lambda\phi^4$ interaction in $d$ dimensions, specializing to $d = 4$ later in the calculation. Since no one-point correlation is generated, we can directly start with the two-point correlation in eq.\ (\ref{eq:2ptdef}). The one-loop diagram that contributes at ${\cal O}(\lambda)$ and the $\delta_m$ and $\delta_r$ counterterm diagrams are shown in fig.\ \ref{fig:2p1l}. Their contribution to the Feynman Green's function is
\bea
    G_{F,{\rm 1-loop}}(k) & = & -\frac{i\lambda}{2} \int_p G_F(k) G_F(p) G_F(k) - i \( \delta_m + k^2 \delta_r \) G_F(k) G_F(k) \, .
\label{eq:phi41loop}
\eea
To calculate this, we only need to find the loop integral over $p$, which is exactly the same as that in eq. (\ref{eq:tp}) for the tadpole contribution to the one-point function in the cubic theory. As found there, the loop integral vanishes in the massless limit that we are working in, and so there is no one-loop contribution to the two-point correlation.

\begin{figure}[!t]
\begin{center}
	\includegraphics[scale=0.6]{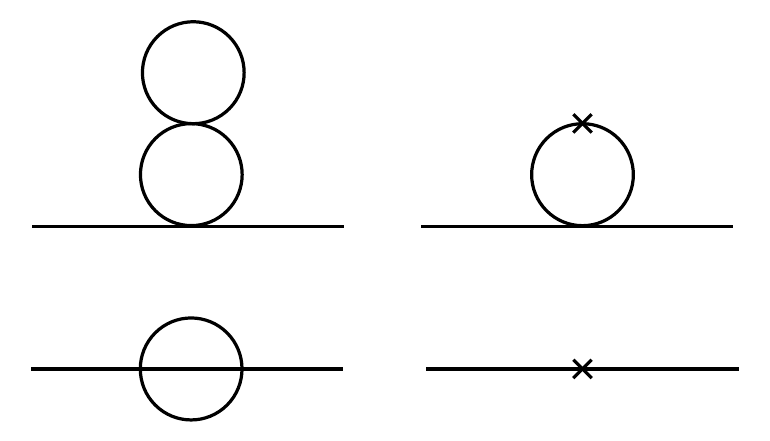}
	\caption{(Left) Connected two-loop contributions to the two-point correlation function in a $\lambda\phi^4$ theory. (Right) The corresponding $\delta_m$ and/or $\delta_r$ counterterm diagrams.}
\label{fig:2p2l}
\end{center}
\end{figure}

To find a non-zero contribution to the two-point correlation, we thus need to calculate the next-order correction. The two-loop diagrams that contribute at ${\cal O}(\lambda^2)$ and the $\delta_m$ and $\delta_r$ counterterm diagrams are shown in fig.\ \ref{fig:2p2l}. We will refer to the top left and bottom left diagrams as the snowman (`sm') and sunset (`ss') diagrams, respectively. The contribution from the snowman diagram and the corresponding counterterm diagrams is
\bea
    G_{F,{\rm 2-loop, sm}}(k) & = & -\frac{\lambda^2}{2} \int_p \int_{p'} G_F(k) G_F(p) G_F(p') G_F(k) \nn \\
    & & \quad - \ \frac{\lambda}{2} \( \delta_m + k^2 \delta_r \) \int_p G_F(k) G_F(p) G_F(k) \, .
\eea
We see that this is proportional to the one-loop contribution in eq.\ (\ref{eq:phi41loop}), and, therefore, also vanishes in the massless limit. The snowman diagram thus does not contribute to the correlation function either.

Let us finally consider the contribution from the sunset diagram and its corresponding counterterm diagrams,
\bea
    G_{F,{\rm 2-loop, ss}}(k) & = & -\frac{\lambda^2}{6} \int_{p} \int_{p'} G_F(k) G_F(p) G_F(p') G_F(k-p-p') G_F(k) \nn \\
    & & \quad - \ i \( \delta_m + k^2 \delta_r \) G_F(k) G_F(k) \, .
\label{eq:phi42loopss}
\eea
Now we only need to find the loop integral over $p$ and $p'$,
\bea
    I_{\lambda\phi^4}(k) & = & \frac{\lambda^2}{6} \int_p\int_{p'} G_F(p)G_F(p')G_F(k-p-p') \, ,
\label{eq:sunsetloop}
\eea
which is again a standard calculation, but is shown using dimensional regularization in appendix\ \ref{app:42l}. Inserting the result from eq.\ (\ref{eq:phi4loop4d}) into eq.\ (\ref{eq:phi42loopss}) gives
\bea
    G_{F,{\rm 2-loop, ss}}(k) & = & G_F^2(k) \[ \frac{i \lambda^2 k^2}{3072 \pi ^4 \epsilon} - \frac{i \lambda^2 k^2}{3072 \pi^4} \left\{ \frac{13}{4} - \gamma + \log \( \frac{4\pi\mu^2}{k^2} \) \right\} - i \( \delta_m + k^2 \delta_r \) \] . \nn \\
\label{eq:phi42loopssp}
\eea
We can cancel the UV divergence with the following choice of counterterms,
\bea
    \delta_m \ = \ 0 \, , \quad \delta_r \ = \ \frac{\lambda^2}{3072 \pi ^4} \[ \frac{1}{\epsilon} + \gamma - \ln (4\pi) \] ,
\label{eq:phi4ct}
\eea
where, as before, we have used the $\overline{\rm MS}$ renormalization scheme. The two-loop correction is then given by the remaining terms in eq.\ (\ref{eq:phi42loopssp}), which we next plug into eq.\ (\ref{eq:ftfinal}), choosing $t>t'$. Taking the limits $k (t-t') \gg 1$ and $\mu (t-t') \gg 1$ gives us a final expression for the connected two-point correlation at two-loops and in the late-time limit,
\bea
    \big\la \hPhi_{\vk}(t) \hPhi_{\vk'}(t') \big\ra_c & = & (2\pi)^3 \delta^3 \big( \vk+\vk' \big) \frac{e^{-i |\vk| (t-t')}}{2|\vk|} \[ 1 - \frac{\lambda^2}{3072\pi^4} \ln\left\{ \mu(t-t') \right\} \] ,
\label{eq:phi42pft4d}
\eea
where we again find a logarithmic divergence as $\mu(t-t') \rightarrow \infty$. Note the similarity between this result and what we found for $\lambda\phi^3$ in 6D, where $\lambda$ was also dimensionless, in eq.\ (\ref{eq:2pft6d}). The secular term here will also exponentiate into a polynomial decay with the WW resummation method that we discuss next.


\section{The Weisskopf-Wigner (WW) resummation method}
\label{sec:ww}

We start this section with an overview of the WW resummation method. Consider a quantum system that is described by a Hamiltonian of the form $\hat{H} = \hat{H}_0 + \hat{H}_i$, where $\hat{H}_0$ is the free part and $\hat{H}_i$ the interaction part. The Hamiltonian can, in general, have an explicit time-dependence as well, but a time-independent Hamiltonian will suffice for the problem that we consider in this paper. Suppose that the system is initialized in the state $|\psi(t_0)\rangle$ at the initial time $t_0$ and is described by the state $|\psi(t)\rangle_I$ in interaction picture at a later time $t$. In terms of the unitary time evolution operator introduced earlier, the time-evolved state is given by $|\psi(t)\rangle_I = \hat{U}_{i,I}(t,t_0) |\psi(t_0)\rangle$. It can also be written in terms of orthonormal basis states $\{ |n\rangle \}$ as
\bea
    |\psi(t)\rangle_I & = & \sum_{n} C_n(t) |n\rangle \, .
\label{stateexpan}
\eea
For the quantum field theories that we consider, $\{ |n\rangle \}$ would be multi-particle Fock states, with the index $n$ referring to all possible momentum configurations in an $n$-particle state. The unitarity of quantum mechanics requires that the probability amplitudes $C_n(t)$ satisfy $\sum_n |C_n(t)|^2 = 1$, when summing over all states in the Hilbert space.

Inserting eq.\ (\ref{stateexpan}) into the Schr\"{o}dinger equation yields a set of coupled first order differential equations for $C_n(t)$, 
\bea
    \dot{C}_n(t) & = & -i \sum_{m} C_m(t) \big\la n \big| \hat{H}_{i,I}(t) \big| m \big\ra \, .
\label{ceq1}
\eea
These equations describe the time evolution of any quantum state in a non-perturbative manner. However, they are typically not solvable for general systems and interactions. The WW method makes two approximations to close this set of equations within a truncated Hilbert space \cite{Boyanovsky:2011xn}: (i) restrict the calculation to second order in perturbation theory and (ii) make a Markov approximation, as we explain further below. The net result of these two approximations is that unitarity is preserved within the truncated Hilbert space or, in other words, the system now lies solely within the truncated Hilbert space. This was shown in, for example, \cite{Lello:2013qza}, and we refer the reader to this paper for a proof. The resulting `resummation' of $C_n(t)$ makes this method particularly useful to study the stability/decay of states in the presence of an interaction, and we will adapt it to calculate resummed (late-time) correlation functions instead. It is also interesting to note that the conservation of probability in the WW method is analogous to ${\rm Tr} \[ \hat{\rho}(t) \] = 1$ for the reduced density matrix $\hat{\rho}(t)$ of an open quantum system that evolves with a master equation.

Suppose the system is in one of the basis states $|A\rangle$ at the initial time $t_0$, so that $C_A(t_0) = 1$ and $C_{n \ne A}(t_0) = 0$. Let us also assume that $|A\rangle$ is connected at first order in perturbation theory to a subset of basis states $\{ |\kappa\rangle \}$ different from $|A\ra$, that is, $\big\la \kappa \big| \hat{H}_{i,I}(t) \big| A \big\ra \ne 0$, with $\langle \kappa | A \rangle = 0$. Then, for the states $| A \rangle$ and $\{ |\kappa\rangle \}$, the coupled equations in eq.\ (\ref{ceq1}) become
\bea
    \dot{C}_A(t) & = & -i C_A(t) \big\la A \big| \hat{H}_{i,I}(t) \big| A \big\ra - i \sum_{\{ \kappa \}} C_{\kappa}(t) \big\la A \big| \hat{H}_{i,I}(t) \big| \kappa \big\ra \, ,
\label{eq:cadotp} \\
    \dot{C}_{\kappa}(t) & = & -i \sum_{m} C_m(t) \big\la \kappa \big| \hat{H}_{i,I}(t) \big| m \big\ra \, ,
\label{eq:ckappadotp}
\eea
where we have separated out the diagonal matrix element or local contribution to $\dot{C}_A(t)$ \cite{Boyanovsky:2012qs}. To solve these equations analytically, we assume that $C_A(t)$ remains close to unity and all other coefficients remain close to zero at all times. This is certainly the case for times close to the initial time $t_0$, but may be violated at later times. Nonetheless, we proceed with this assumption. Then, on keeping only the leading order term that is proportional to $C_A(t)$ on the right hand side of eq.\ (\ref{eq:ckappadotp}), it simplifies to
\bea
    \dot{C}_{\kappa}(t) & \approx & -i C_A(t) \big\la \kappa \big| \hat{H}_{i,I}(t) \big| A \big\ra \, .
\label{eq:ckappadot}
\eea
We can now integrate this equation and substitute the result back into eq.\ (\ref{eq:cadotp}) to get
\bea
    \dot{C}_A(t) & = & -i C_A(t) \big\la A \big| \hat{H}_{i,I}(t) \big| A \big\ra - \int_{t_0}^t \d t_2 \, \Sigma_A(t,t_2) C_A(t_2) \, ,
\label{eq:cadot}
\eea
where we have defined the retarded self-energy,
\bea
    \Sigma_A(t,t_2) & = & \sum_{\{ \kappa \}} \big\la A \big| \hat{H}_{i,I}(t) \big| \kappa \big\ra \big\la \kappa \big| \hat{H}_{i,I}(t_2) \big| A \big\ra \, ,
\label{eq:sigma}
\eea
for $t > t_2$. Eq.\ (\ref{eq:cadot}) suggests that $C_A(t)$ evolves slowly in time since its time derivative is proportional to the interaction Hamiltonian, which is consistent with our previous assumption that it remains close to unity at all times. We will thus take $C_A(t_2)$ out of the integral in eq.\ (\ref{eq:cadot}), evaluating it at the upper limit of the integral $t$. This Markov approximation essentially erases the memory of $C_A(t)$ and is also why the WW method is only trustworthy at late times. We can now solve for $C_A(t)$ to obtain a final expression for it,
\bea
    C_A(t) & = & {\rm Exp} \[ -i \int_{t_0}^t \d t_1 \, \big\la A \big| \hat{H}_{i,I}(t_1) \big| A \big\ra - \int_{t_0}^t \d t_1 \int_{t_0}^{t_1} \d t_2 \, \Sigma_A(t_1,t_2) \] .
\label{eq:casolww}
\eea
In the notation of \cite{Boyanovsky:2011xn,Boyanovsky:2012qs}, the exponent above can be denoted as $-\int_{t_0}^t \d t_1 \, W_0(t_1,t_1)$, with $W_0(t_1,t_1) = i \big\la A \big| \hat{H}_{i,I}(t_1) \big| A \big\ra + \int_{t_0}^{t_1} \d t_2 \, \Sigma_A(t_1,t_2)$. Lastly, once we have $C_A(t)$, we can use it in eq.\ (\ref{eq:ckappadot}) to solve for the remaining coefficients $C_{\kappa}(t)$.

\subsubsection*{Correlation functions of a quantum field}

Before discussing the calculation of correlation functions, we will adapt the above analysis to a quantum field. We are interested in the canonically quantized field $\hat{\Phi}_I(t,\vx) = \int_{\vk} e^{i\vk\cdot\vx} \hat{\Phi}_{\vk,I}(t)$, with
\bea
	\hPhi_{\vk,I}(t) & = & f_k^>(t) \hat{a}_{\vk} + f_k^<(t) \hat{a}_{-\vk}^{\dagger} \, ,
\label{eq:cq}
\eea
where $\hat{a}_{\vk}$ and $\hat{a}_{\vk}^{\dagger}$ are Schr\"odinger picture ladder operators defined at the initial time $t_0$ and we write expressions for the mode functions $f_k^{\gtrless}(t)$ later. We also choose the following normalization of states: $\hat{a}_{\vk}^{\dagger} | 0 \ra = | \vk \ra$, $\la \vk | \vk' \ra = (2\pi)^{d-1} \delta^{d-1} \big( \vk-\vk' \big)$, and $\big[\hat{a}_{\vk}, \hat{a}_{\vk'}^{\dagger}\big] = (2\pi)^{d-1} \delta^{d-1} \big( \vk-\vk' \big)$.\footnote{We can alternatively work with the box-normalized field and convert from (dimensionless) box-normalized quantities to their continuous space analogs by replacing $\frac{1}{V} \sum_{\vk} \rightarrow \int_{\vk}$, $V \delta_{\vk, \vk'} \rightarrow (2\pi)^{d-1} \delta^{d-1} \big( \vk - \vk' \big)$, and $\sqrt{V} \hat{a}_{\vk} \rightarrow \hat{a}_{\vk}$, where $V = \int \d^{d-1}x = (2\pi)^{d-1} \delta^{d-1} \big( \vec{0} \big)$ is the volume. This is useful for keeping track of volume factors so that dimensions are consistent in the continuum case.} The state $C_n(t) |n\rangle$ in eq.\ (\ref{stateexpan}) is now the $n$-particle state,
\bea
    C_n(t) |n\rangle & = & \frac{1}{n!} \int_{\vk_1} \ldots \int_{\vk_n} C_{\vk_1, \ldots, \vk_n}(t) \big| 1_{\vk_1}, \ldots, 1_{\vk_n} \big\rangle \, ,
\label{eq:Fockexp}
\eea
and we will continue to use the shorthand $C_n(t) |n\rangle$ to denote this state. Suppose the field is in the state $|A\rangle$ at the initial time $t_0$ as before. Then, using the expansion in eq. (\ref{eq:Fockexp}) and following the steps outlined in eqs.\ (\ref{eq:cadotp})-(\ref{eq:casolww}), eq.\ (\ref{eq:casolww}) becomes
\bea
    C_A(t) & = & {\rm Exp} \[ -\frac{i}{V^{n_A}} \int_{t_0}^t \d t_1 \, \big\la A \big| \hat{H}_{i,I}(t_1) \big| A \big\ra - \int_{t_0}^t \d t_1 \int_{t_0}^{t_1} \d t_2 \, \Sigma_A(t_1,t_2) \] ,
\label{eq:casolft}
\eea
with
\bea
    \Sigma_A(t_1,t_2) & = & \sum_{ \{ \kappa \}} \frac{1}{\kappa!V^{n_A}} \int_{\vk_1} \ldots \int_{\vk_{\kappa}} \big\la A \big| \hat{H}_{i,I}(t_1) \big|  1_{\vk_1}, \ldots, 1_{\vk_\kappa} \big\ra \big\la 1_{\vk_1}, \ldots, 1_{\vk_\kappa} \big| \hat{H}_{i,I}(t_2) \big| A \big\ra \, , \nn \\
   \label{eq:Sigmaft}
\eea
where $n_A$ is the number of particles in the initial state. Once we have $C_A(t)$, we can again use it to calculate the remaining coefficients $C_{\kappa}(t)$.

With the resummed probability amplitudes in hand, it is now straightforward to obtain resummed correlation functions for a quantum field. Let us first consider its one-point function calculated in the initial state $\hat{\rho}(t_0) = |0\ra \la 0|$,
\bea
    \big\la \hPhi_{\vk}(t) \big\ra & = & \big\la 0\big|\hat{U}^{\dagger}_{i,I}(t,t_0) \hPhi_{\vk,I}(t) \hat{U}_{i,I}(t,t_0) \big|0\big\ra \, .
\label{eq:ww1p}
\eea
We will denote the time-evolved state $\hat{U}_{i,I}(t,t_0) |0\ra$ here as $|\Omega(t)\ra = \sum_n C_n(t) |n\ra$, as we did in eq.\ (\ref{stateexpan}). For a given interaction Hamiltonian and in the limit of $t \gg t_0$, the WW method gives us resummed expressions for $C_n(t)$ and thus $|\Omega(t)\ra$. The resummed one-point function then becomes $\big\la\Omega(t)\big| \hPhi_{\vk,I}(t) \big|\Omega(t)\big\ra$, which can be found by plugging in eq. (\ref{eq:cq}).

Let us next consider the two-point correlation,
\bea
    \big\la \hPhi_{\vk}(t) \hPhi_{\vk'}(t') \big\ra_c & = & \big\la 0\big| \hat{U}^{\dagger}_{i,I}(t,t_0) \hPhi_{\vk,I}(t) \hat{U}_{i,I}(t,t') \hPhi_{\vk',I}(t') \hat{U}_{i,I}(t',t_0) \big|0\big\ra_c \, ,
\label{eq:ww2p}
\eea
for $t > t'$. We can now obtain the resummed result by applying the WW method twice. First, in the limit of $t' \gg t_0$, we can replace $\hat{U}_{i,I}(t',t_0) |0\ra$ with $|\Omega(t')\rangle$. Second, we can find $\hPhi_{\vk,I}(t') |\Omega(t')\rangle$ and, taking the resulting state as the new initial state (at time $t'$), apply the WW method again to find the action of $\hat{U}_{i,I}(t,t')$ on it in the limit of $t \gg t'$. The overlap of the final state with $\langle \Omega(t) | \hPhi_{\vk,I}(t)$ gives us the resummed late-time two-point correlation.
   
We use these expressions to calculate late-time correlations for the Lagrangian densities in eqs.\ (\ref{eq:lag3}) and (\ref{eq:lag4}) in the two subsections below.

\subsection{\texorpdfstring{$\lambda\phi^3$}{lp3} in 4D and 6D}

The Hamiltonian for the $\lambda\phi^3$ theory can be obtained by taking a Legendre transform of the Lagrangian density in eq.\ (\ref{eq:lag3}) and integrating over space. We first need the momentum conjugate to $\phi$, given by $\pi = \partial {\cal L}/\partial \dot{\phi} = Z_r \dot{\phi}$, where we define $Z_r = 1 + \delta_r$. Denoting the corresponding operator $\hat{\Pi}(\vx)$ and setting the mass of the field to zero, the Hamiltonian becomes
\bea
    \hat{H} & = & \int \d^d x \[ \frac{1}{2Z_r} \hat{\Pi}^2 + \frac{Z_r}{2} (\partial_i \hat{\Phi})^2 + \frac{1}{2} \delta_m \hat{\Phi}^2 + \frac{1}{3!} \lambda \hat{\Phi}^3 - Y\hat{\Phi} \] .
\label{eq:ham3}
\eea
We will refer to the last three terms above as the interaction Hamiltonian $\hat{H}_i$ and switch to Fourier space as before. We can now write the interaction picture field in terms of ladder operators as in eq.\ (\ref{eq:cq}), with the mode functions given by
\bea
    f_k^>(t) & = & \frac{e^{-i |\vk| (t-t_0)}}{\sqrt{2Z_r |\vk|}}
\label{eq:fg}
\eea
and $f_k^<(t) = f_k^{>*}(t)$. One way to obtain the mode functions would be to first define a canonically normalized field $\chi = \sqrt{Z_r} \phi$, whose solution is the usual one for a Klein-Gordon field, and then relate the solution back to $\phi$. The resulting factor of $\sqrt{Z_r}$ in the denominator in eq.\ (\ref{eq:fg}) is important for the resummation, as we will see later.

With the $\lambda\phi^3$ Hamiltonian in hand, we first want to ensure that the one-point function vanishes, as we did for the perturbative calculation. The set of states that are connected to the initial state $|0\rangle$ through this interaction Hamiltonian at first order in perturbation theory is $\{ |1\rangle, |3\rangle \}$. The WW method then tells us that in the limit of $t \gg t_0$, the resummed state $\hat{U}_{i,I}(t,t_0) |0\ra$ is given by $|\Omega(t) \rangle = C_0(t) |0\rangle + C_1(t) |1\rangle + C_3(t) |3\rangle$. The one-point function in eq.\ (\ref{eq:ww1p}) thus becomes $\big\la \hPhi_{\vk}(t) \big\ra = C_0^*(t) C_{\vk}(t) f_k^>(t) + C_{\vk}^*(t) C_0(t) f_k^<(t)$. This vanishes if we set $C_{\vk}(t) = 0$ at all times which, from eq.\ (\ref{eq:ckappadot}), translates into the condition that $\big\la 1_{\vk} \big| \hat{H}_{i,I}(t) \big| 0 \big\ra = 0$. We can obtain this matrix element by plugging in the interaction Hamiltonian from eq.\ (\ref{eq:ham3}) (converted to Fourier space) and using eq.\ (\ref{eq:cq}) for $\hPhi_{\vk,I}(t)$,
\bea
    \big\la 1_{\vk} \big| \hat{H}_{i,I}(t) \big| 0 \big\ra & = & (2\pi)^{d-1} \delta^{d-1}\big( \vk \big) \[ \frac{\lambda}{2} \int_{\vp} f_k^<(t) f_p^>(t) f_p^<(t) - Y f_k^<(t) \] .
\label{eq:wwy}
\eea
Setting this to zero yields
\bea
    Y & = & \frac{\lambda}{2} \int_{\vp} f_p^>(t) f_p^<(t) \, ,
\eea
which matches what we found in eq. (\ref{eq:linct}).\footnote{We have used $\int_p G_F(p) = \lim_{t'\to t} \int_p \frac{i e^{-ip_0(t-t')}}{p_0^2-|\vec{p}|^2+i\epsilon} = \int_{\vp} f_p^>(t) f_p^<(t)$.} As noted before, adding the $Y$ counterterm is equivalent to normal ordering the cubic interaction. We will take advantage of this fact when finding non-vanishing matrix elements of the interaction Hamiltonian below.

We next want to find the two-point correlation using the WW method twice as mentioned after eq.\ (\ref{eq:ww2p}). In the limit of $t' \gg t_0$, we can replace $\hat{U}_{i,I}(t',t_0) |0\ra$ with $|\Omega(t')\rangle$ which, for the normal ordered interaction Hamiltonian $\lambda {\,:}\hat{\Phi}^3{:\,}$, can now be written as $|\Omega(t') \rangle = C_0(t') |0\rangle + C_3(t') |3\rangle$. We show in appendix\ \ref{app:C0} that $C_0(t')$ here is simply a phase factor in both 4D and 6D, and is in fact unity in the late-time limit that we are interested in studying. The (free theory) vacuum is, therefore, stable under a $\lambda\phi^3$ interaction in Minkowski. The two-point correlation can then be written in the limit of $t, t' \gg t_0$ as
\bea
    \big\la \hPhi_{\vk}(t)\hPhi_{\vk'}(t') \big\ra_c & = & C_0^*(t) C_0(t') \big\la 0\big| \hPhi_{\vk,I}(t) \hat{U}_{i,I}(t,t') \hPhi_{\vk',I}(t') \big|0\big\ra_c \, ,
\label{eq:ww2pc0}
\eea
which becomes
\bea
    \big\la \hPhi_{\vk}(t)\hPhi_{\vk'}(t') \big\ra_c & = & f_k^>(t) f_{k'}^<(t') \big\la 1_{\vk} \big| \hat{U}_{i,I}(t,t') \big| 1_{\vk'} \big\ra_c
\label{eq:ww2p2}
\eea
on setting $C_0(t)$ to unity and using eq.\ (\ref{eq:cq}) for $\hPhi_{\vk,I}(t)$. We can now apply the WW method again to write $\hat{U}_{i,I}(t,t') \big| 1_{\vk'} \big\ra$ in a resummed form. For a $\lambda {\,:}\hat{\Phi}^3{:\,}$ interaction, the set of states connected to the initial state $|0\ra$ at first order in perturbation theory is $\{ |3\ra \}$ while that connected to $|1\ra$ is $\{ |2\ra, |4\ra \}$. We can thus write $\hat{U}_{i,I}(t,t') |1_{\vk'}\ra = C_{\vk'}(t,t') |1_{\vk'}\ra + C_{2,{\vk'}}(t,t') |2_{\vk'}\ra + C_{4,{\vk'}}(t,t') |4_{\vk'}\ra$ for $t \gg t'$, where the subscripts ${\vk'}$ indicate the total momentum of any state and we write $C_n$ as a function of two times since the initial time is now $t'$ rather than $t_0$. What appears in eq.\ (\ref{eq:ww2p2}) is the overlap of the final state with $\big\la 1_{\vk} \big|$, and so we can simply replace $\big\la 1_{\vk} \big| \hat{U}_{i,I}(t,t') \big| 1_{\vk'} \big\ra_c$ there with $(2\pi)^{d-1} \delta^{d-1} \big( \vk+\vk' \big) C_{\vk'}(t,t')$. Thus we have so far that
\bea
    \big\la \hPhi_{\vk}(t)\hPhi_{\vk'}(t') \big\ra_c & = & (2\pi)^{d-1} \delta^{d-1} \big( \vk+\vk' \big) f_k^>(t) f_{k'}^<(t') C_{\vk'}(t,t') \, ,
\label{eq:ww2p3}
\eea
and we are only left to calculate $C_{\vk'}(t,t')$.

Let us now calculate $C_{\vk'}(t,t')$, starting in the state $\big| 1_{\vk'} \big\ra$ at time $t'$. From eq.\ (\ref{eq:casolft}), we have
\bea
    C_{\vk'}(t,t') = {\rm Exp} \[ -\frac{i}{V} \int_{t'}^t \d t_1 \, \big\la 1_{\vk'} \big| \hat{H}_{i,I}(t_1) \big| 1_{\vk'} \big\ra - \int_{t'}^t \d t_1 \int_{t'}^{t_1} \d t_2 \, \Sigma_{\vk'}(t_1,t_2) \] .
\label{eq:cksolww}
\eea
The contribution from the first term in the exponential vanishes for $\lambda {\,:}\hat{\Phi}^3{:\,}$ but is non-zero for the counterterms, which we return to later in the calculation. The self energy, given by eq.\ (\ref{eq:Sigmaft}), itself has two contributions, one that corresponds to vacuum diagrams and one to connected diagrams. The vacuum one, that we can denote $\Sigma_0(t_1,t_2)$, leads to similar factors as $C_0(t')$ found earlier, and again goes to the unity in the $t \gg t'$ limit where we can trust the WW approximation. The connected contribution to the self energy, on the other hand, is given by
\bea
    \Sigma_{\vk',c}(t_1,t_2) & = & \frac{1}{2V} \int_{\vp_1} \int_{\vp_2} \big\la 1_{\vk'} \big| \hat{H}_{i,I}(t_1) \big| 1_{\vp_1}, 1_{\vp_2} \big\ra \big\la 1_{\vp_1}, 1_{\vp_2} \big| \hat{H}_{i,I}(t_2) \big| 1_{\vk'} \big\ra
\label{sigma1c} \, ,
\eea
which, on using eq.\ (\ref{eq:cq}) for $\hPhi_{\vk,I}(t)$ in the interaction Hamiltonian, becomes
\bea
    \Sigma_{\vk',c}(t_1,t_2) & = & \frac{\lambda^2}{2} f_{k'}^>(t_2) f_{k'}^<(t_1) \int_{\vp} f_p^>(t_1) f_p^<(t_2) f_{|\vk'-\vp|}^>(t_1) f_{|\vk'-\vp|}^<(t_2) \nn \\
    & = & \frac{\lambda^2}{2} \frac{e^{i|\vk'|(t_1-t_2)}}{2|\vk'|} \int_{\vp} \frac{e^{-i|\vp|(t_1-t_2)}}{2|\vp|} \frac{e^{-i|\vk'-\vp|(t_1-t_2)}}{2|\vk'-\vp|} \, ,
\label{sigma1cp}
\eea
where we have set $Z_r$ to unity in the denominator of the mode function at this order and the volume factors have canceled out. What appears in $C_{\vk'}(t,t')$ is a double time integral $\int \d t_1 \int \d t_2$ over $\Sigma_{\vk'}(t_1,t_2)$. Because of the underlying time-translation invariance, we should be able to express our results in terms of time differences and thus change the time integrals to $\int \d \tau \int \d \Delta$, with $\tau = t_1 - t'$ and $\Delta = t_1 - t_2$. With this change of variables, eq.\ (\ref{sigma1cp}) becomes
\bea
    \Sigma_{\vk',c}(\tau,\Delta) & = & \frac{\lambda^2}{2^{d+2} \pi^{d/2} \Gamma(d/2-1)} \frac{e^{i|\vk'| \Delta}}{|\vk'|^2} \int_0^{\infty} \d|\vp| \int_{||\vk'| - |\vec{p}||}^{|\vk'| + |\vec{p}|} \d |\vq| \, |\vp|^{d-4} \nn \\
    & & \qquad \times \[ 1 - \frac{\(|\vk'|^2 + |\vp|^2 - |\vq|^2 \)^2}{4 |\vk'|^2 |\vp|^2} \]^{d/2-2} e^{-i(|\vp|+|\vq|) \Delta} \, .
\label{sigma1cp2}
\eea
where we have used $\int \d^{d} x = \frac{2\pi^{(d-1)/2}}{\Gamma [(d-1)/2]} \int_0^{\infty} \d x \int_{-1}^1 \d \cos\theta \, x^{d-1} \sin^{d-3}\theta$, defined $\vq = \vk'-\vp$, so that $|\vq| = \( |\vk'|^2 + |\vp|^2 - 2|\vk'| |\vp| \cos\theta \)^{1/2}$, and changed the integral over $\cos\theta$ to $|\vq|$ using $\d \cos\theta = -\frac{|\vq|}{|\vk'| |\vp|} \, \d |\vq|$.

We will now specialize to either $d=4$ or $d=6$ to obtain $\int_0^{t-t'} \d \tau \int_0^{\tau} \d \Delta \, \Sigma_{\vk,c}(\tau,\Delta)$ and thus $C_{\vk'}(t,t')$.

\subsubsection*{(i) $\lambda\phi^3$ in 4D}

Let us first set $d = 4 + \epsilon$. The loop integral in eq.\ (\ref{sigma1cp2}) does not need a regulator if we convert $\Delta$ into a Euclidean time coordinate. We thus define $\Delta = -i\bar{\Delta}$ and use dimensional regularization on $\bar{\Delta}$ \cite{Chaykov:2022pwd}. In other words, we replace $\int \d \bar{\Delta}$ with $\mu^{\epsilon} \int \d^{1+\epsilon} \bar{\Delta} = \mu^{\epsilon} \frac{\pi^{(1+\epsilon)/2}}{\Gamma[(1+\epsilon)/2]} \int \d \bar{\Delta} \, \bar{\Delta}^{\epsilon}$ where, as before, $\mu$ is a parameter with dimensions of mass, and set $d = 4$ in eq.\ (\ref{sigma1cp2}). $\Sigma_{\vk',c}(\tau,\Delta)$ then becomes\footnote{We did not cancel out the $e^{\pm|\vk'|\bar{\Delta}}$ factors in eq.\ (\ref{eq:Sigmaphi34d}) simply to make the connection with the corresponding equation in our companion paper \cite{Chaykov:2022pwd} clearer.}
\bea
    \Sigma_{\vk',c}(\tau,\Delta) & = & \frac{\lambda^2}{64 \pi^2} \frac{e^{|\vk'| \bar{\Delta}}}{|\vk'|^2} \int_0^{\infty} \d|\vp| \int_{||\vk'| - |\vec{p}||}^{|\vk'| + |\vec{p}|} \d |\vq| \, e^{-(|\vp|+|\vq|) \bar{\Delta}} \nn \\
    & = & \( \frac{e^{|\vk'| \bar{\Delta}}}{2|\vk'|} \) \frac{\lambda^2}{2} \frac{e^{-|\vk'|\bar{\Delta}}}{16 \pi^2 \bar{\Delta}} \, ,
\label{eq:Sigmaphi34d}
\eea
and its double time integral gives
\bea
    & & \int \d \tau \int \d \Delta \, \Sigma_{\vk',c}(\tau,\Delta) \ = \ -\frac{i\lambda^2 \mu^{\epsilon}}{64 \pi^2 |\vk'|} \frac{\pi^{(1+\epsilon)/2}}{\Gamma[(1+\epsilon)/2]} \int_0^{t-t'} \d \tau \int_0^{i\tau} \d \bar{\Delta} \, \bar{\Delta}^{\epsilon-1} \nn \\
    & & \qquad \qquad \quad = -\frac{i\lambda^2 (t-t')}{64 \pi^2 |\vk'| \epsilon} - \frac{i \lambda^2 (t-t')}{128 \pi^2 |\vk'|} \[i\pi + \ln \left\{ 4\pi\mu^2 (t-t')^2 \right\} - 2 + \gamma \] ,
\label{eq:c1w1piece4d}
\eea
after series expanding in $\epsilon$ in the second equality. The $\epsilon \rightarrow 0$ divergence can be canceled by appropriately choosing the $\delta_m$ and $\delta_r$ counterterms that enter at leading order through the diagonal contribution in eq.\ (\ref{eq:cksolww}) and the $f_k^>(t) f_{k'}^<(t')$ term in eq.\ (\ref{eq:ww2p3}), respectively,
\bea
    i \int_{t'}^t \d t_1 \, \big\la 1_{\vk'} \big| \hat{H}_{i,I}(t_1) \big| 1_{\vk'} \big\ra & = & \frac{i}{2 |\vk'|} \delta_m (t-t') \, ,
\label{eq:wwctdeltam} \\
    f_k^>(t) f_{k'}^<(t') \delta^{d-1} \big( \vk+\vk' \big) & = & \frac{1}{2 (1+\delta_r) |\vk|} e^{-i |\vk| (t-t')} \delta^{d-1} \big( \vk+\vk' \big) \, .
\label{eq:wwctdeltar}
\eea
We see that the following choice of counterterms cancels the $\epsilon \rightarrow 0$ divergence in eq.\ (\ref{eq:c1w1piece4d}),
\bea
    \delta_m \ = \ \frac{\lambda^2}{64 \pi^2} \[ \frac{2}{\epsilon} + \gamma + \ln (4\pi) \] , \quad \delta_r \ = \ 0 \, ,
\label{eq:d4ctww}
\eea
where, as before, we have used the $\overline{\rm MS}$ renormalization scheme. Note that the counterterms do not have to coincide with those found using the in-out procedure in eq.\ (\ref{eq:d4ct}) since the regularization schemes in the two calculations are not identical. Plugging the remaining terms from eq.\ (\ref{eq:c1w1piece4d}) into eq.\ (\ref{eq:cksolww}) and ultimately into eq.\ (\ref{eq:ww2p3}) gives us the final connected two-point correlation at one-loop,
\bea
    \big\la \hPhi_{\vk}(t) \hPhi_{\vk'}(t') \big\ra_c & = & (2\pi)^3 \delta^3 \big( \vk+\vk' \big) \frac{e^{-i |\vk| (t-t')}}{2|\vk|} \, {\rm Exp} \[ - \frac{\lambda^2(t-t')}{64\pi^2|\vk|} \( \frac{\pi}{2} - i \ln \left\{ \mu (t-t') \right\} + i \) \] . \nn \\
\label{eq:ww2pact4d}
\eea
The decay here is an exact exponential of the real part of the secular divergences that we found in eq.\ (\ref{eq:2pft4d}). The phase factors are slightly different, however, with the main difference being a term proportional to $\ln \big( |\vk|/\mu \big)$, that can also be attributed to regularization details, since we regulate the spatial part of the momentum integral as well in the in-out calculation, unlike here. The correlator no longer diverges as $\lambda^2 (t-t')/|\vk| \rightarrow \infty$ or $\mu (t-t') \rightarrow \infty$ and instead decays exponentially, suggesting a successful late-time resummation of the perturbative result.

\subsubsection*{(ii) $\lambda\phi^3$ in 6D}

Let us next set $d = 6+\epsilon$. As in the 4D case, we define $\Delta = -i\bar{\Delta}$ and regulate the $\bar{\Delta}$ integral, setting $d=6$ in eq.\ (\ref{sigma1cp2}). $\Sigma_{\vk',c}(\tau,\Delta)$ now becomes
\bea
    \Sigma_{\vk',c}(\tau,\Delta) & = & \frac{\lambda^2}{256 \pi^3} \frac{e^{|\vk'| \bar{\Delta}}}{|\vk'|^2} \int_0^{\infty} \d|\vp| \int_{||\vk'| - |\vec{p}||}^{|\vk'| + |\vec{p}|} \d |\vq| \[ |\vp|^2 - \frac{\(|\vk'|^2 + |\vp|^2 - |\vq|^2 \)^2}{4 |\vk'|^2} \] e^{-(|\vp|+|\vq|) \bar{\Delta}} \nn \\
    & = & \( \frac{e^{|\vk'| \bar{\Delta}}}{2|\vk'|} \) \frac{\lambda^2}{2} \frac{e^{-|\vk'|\bar{\Delta}} (1 + |\vk'| \bar{\Delta}) }{192 \pi^3 \bar{\Delta}^3} \, ,
\label{sigmaafterkphi36d}
\eea
and its double time integral gives
\bea
    \int \d \tau \int \d \Delta \, \Sigma_{\vk',c}(\tau,\Delta) & = & \frac{\lambda^2}{768 \pi^3 \epsilon} + \frac{\lambda^2}{1536 \pi^3} \Bigg[ 2 \ln \left\{\mu (t-t') \right\} + 2 + \gamma + \ln(4\pi) \nn \\
    & & \qquad + \ \frac{i}{|\vk'|}(t-t')^{-1} + i\pi \Bigg] \, ,
\label{eq:c1w1piece6d}
\eea
after series expanding in $\epsilon$ (the integrals converge for $\epsilon > 2$ but we can analytically continue the result to smaller $\epsilon$). The $\delta_m$ and $\delta_r$ counterterms contribute as given in eqs.\ (\ref{eq:wwctdeltam}) and (\ref{eq:wwctdeltar}) and we see that the following choice of counterterms cancels the $\epsilon \rightarrow 0$ divergence above,
\bea
    \delta_m \ = \ 0 \, , \quad \frac{1}{1+\delta_r} \ = \ {\rm Exp} \[ \frac{\lambda^2}{1536 \pi^3 }\left(\frac{2}{\epsilon}+\gamma+\ln(4\pi)\right) \] ,
\label{eq:d4ctww}
\eea
where we have again used the $\overline{\rm MS}$ renormalization scheme. Taylor-expanding the left and right hand sides of the second expression above yields $\delta_r = -\frac{\lambda^2}{1536 \pi^3 } \[ \frac{2}{\epsilon} + \gamma + \ln(4\pi) \]$, following which these counterterms match in form with those found using in-out in eq. (\ref{eq:d6ct}), although we do not expect them to exactly coincide, as mentioned earlier. Substituting the remaining terms from eq.\ (\ref{eq:c1w1piece6d}) into eqs.\ (\ref{eq:ww2p3}) and (\ref{eq:cksolww}) then gives us the final connected two-point correlation at one-loop,
\bea
    \big\la \hPhi_{\vk}(t) \hPhi_{\vk'}(t') \big\ra_c & = & (2\pi)^5 \delta^5 \big( \vk+\vk' \big) \frac{e^{-i |\vk| (t-t')}}{2|\vk|} \, {\rm Exp} \Bigg[ -\frac{\lambda ^2}{768 \pi^3} \nn \\
    & & \qquad \times \, \( \ln \left\{ \mu  (t-t') \right\} + 1 + \frac{i}{2|\vk|} (t-t')^{-1} + \frac{i\pi}{2} \) \Bigg] \, .
\label{eq:ww2pact6d}
\eea
The decay here is an exact exponential of the (real) secular divergence that we found in eq.\ (\ref{eq:2pft6d}). The extra terms are either subdominant in the late-time limit or phase factors. The correlator no longer diverges as $\mu (t-t') \rightarrow \infty$ and instead shows a polynomial decay, again suggesting a successful late-time resummation of the perturbative result.

\subsection{\texorpdfstring{$\lambda\phi^4$}{lp4} in 4D}

We next consider the $\lambda \phi^4$ theory in $d$ dimensions, specializing to 4D later in the calculation. Similar to the $\lambda \phi^3$ case, we first write its Hamiltonian by performing a Legendre transformation on the Lagrangian density in eq.\ (\ref{eq:lag4}), setting the field mass to zero,
\bea
    \hat{H} & = & \int \d^d x \[ \frac{1}{2Z_r} \hat{\Pi}^2 + \frac{Z_r}{2} (\partial_i \hat{\Phi})^2 + \frac{1}{2} \delta_m \hat{\Phi}^2 + \frac{1}{4!} \lambda \hat{\Phi}^4 \] .
\label{eq:ham4}
\eea
Since this interaction leaves the one-point function unchanged (and zero), we can directly focus on the two-point correlator calculation. Further, since the Mikowski vacuum is stable under a $\lambda\phi^4$ interaction \cite{Boyanovsky:2011xn}, we again expect $C_0(t')$ to at most be a time-dependent phase factor. We show in appendix \ref{app:D0} that it is in fact unity in the late-time limit that we are interested in studying, similar to the $\lambda\phi^3$ case. Then, following the same procedure as for $\lambda\phi^3$, the connected two-point correlator is again given by eq.\ (\ref{eq:ww2p3}) with $C_{\vk'}(t,t')$ given by eq.\ (\ref{eq:cksolww}). The contribution from the first term in the exponential vanishes in the massless limit but is non-zero for the counterterms, which we return to later in the calculation. Also, the vacuum contribution to the self-energy goes to unity in the late-time limit, similar to what we found for $\lambda\phi^3$, since $C_0(t')$ goes to unity. The connected contribution to the self-energy, on the other hand, is now given by
\bea
    \Sigma_{\vk',c}(t_1,t_2) & = & \frac{1}{6V} \int_{\vp_1} \int_{\vp_2} \int_{\vp_3} \big\la 1_{\vk'} \big| \hat{H}_{i,I}(t_1) \big| 1_{\vp_1},  1_{\vp_2}, 1_{\vp_3} \big\ra \big\la 1_{\vp_1}, 1_{\vp_2}, 1_{\vp_3} \big| \hat{H}_{i,I}(t_2) \big| 1_{\vk'} \big\ra \nn \\
    & = & \frac{\lambda^2}{6} \frac{e^{i |\vk'|(t_1-t_2)}}{2|\vk'|} \int_{\vp_1} \int_{\vp_2} \frac{e^{-i|\vp_1|(t_1-t_2)}}{2|\vp_1|} \frac{e^{-i|\vp_2|(t_1-t_2)}}{2|\vp_2|} \frac{e^{-i|\vk'-\vp_1-\vp_2|(t_1-t_2)}}{2|\vk'-\vp_1-\vp_2|} \, , \quad
\label{sigma1cpphi4}
\eea
where we have used eq.\ (\ref{eq:cq}) for $\hPhi_{\vk,I}(t)$ in the interaction Hamiltonian, set $Z_r$ to unity in the denominator of the mode function at this order, and the volume factors have again canceled out. On changing the time variables as before and setting the number of spatial dimensions to three, eq.\ (\ref{sigma1cpphi4}) becomes
\bea
    \Sigma_{\vk',c}(\tau,\Delta) & = & \frac{\lambda^2}{1536 \pi^4} \frac{e^{i|\vk'| \Delta}}{|\vk'|^2} \int_0^{\infty} \d |\vp_1| \int_{||\vk'| - |\vec{p}_1||}^{|\vk'| + |\vec{p}_1|} \d |\vq_1| \int_0^{\infty} \d |\vp_2| \int_{||\vq_1| - |\vp_2||}^{|\vq_1| + |\vp_2|} \d |\vq_2| \nn \\
    & & \qquad \times \ e^{-i \( |\vec{p}_1| + |\vec{p}_2| + |\vq_2| \) \Delta} \, ,
\label{sigma1cp2phi4}
\eea
where we have defined $\vq_1 = \vk' - \vp_1$ and $\vq_2 = \vq_1 - \vp_2$, so that $|\vq_1| = \Big( |\vk'|^2 + |\vp_1|^2 - 2|\vk'| |\vp_1| \cos\theta_1 \Big)^{1/2}$ and $|\vq_2| = \Big( |\vq_1|^2 + |\vp_2|^2 - 2|\vq_1| |\vp_2| \cos\theta_2 \Big)^{1/2}$, and changed the integrals over $\cos\theta_1$ and $\cos\theta_2$ to $|\vq_1|$ and $|\vq_2|$ using $\d \cos\theta_1 = -\frac{|\vq_1|}{|\vk'| |\vp_1|} \, \d |\vq_1|$ and $\d \cos\theta_2 = -\frac{|\vq_2|}{|\vq_1| |\vp_2|} \, \d |\vq_2|$, respectively.

We now regulate the loop integrals in eq.\ (\ref{sigma1cp2phi4}) by means of a Euclidean time coordinate, defining $\Delta = -i\bar{\Delta}$, as before. We also set $d = 4+\epsilon$, absorbing the fractional dimension fully into the time coordinate, so that the number of spatial dimensions is still three. $\Sigma_{\vk',c}(\tau,\Delta)$ then becomes
\bea
    \Sigma_{\vk',c}(\tau,\Delta) & = & \( \frac{e^{|\vk'| \bar{\Delta}}}{2|\vk'|} \) \frac{\lambda^2}{6} \frac{e^{-|\vk'|\bar{\Delta}} (1 + |\vk'| \bar{\Delta}) }{256 \pi^4 \bar{\Delta}^3} \, ,
\eea
which is exactly the same (up to a constant) as what we found for $\lambda \phi^3$ in 6D in eq.\ (\ref{sigmaafterkphi36d}), with both interactions sharing the fact that $\lambda$ is dimensionless. The double time integral of $\Sigma_{\vk',c}(\tau,\Delta)$ now gives
\bea
    \int \d \tau \int \d \Delta \, \Sigma_{\vk',c}(\tau,\Delta) & = & \frac{\lambda^2}{3072 \pi^4 \epsilon} + \frac{\lambda^2}{6144 \pi^4} \Bigg[ 2 \ln \left\{ \mu (t-t') \right\} + 2 + \gamma + \ln(4\pi) \nn \\
    & & \qquad + \ \frac{i}{|\vk'|}(t-t')^{-1} + i\pi \Bigg] \, ,
\label{eq:c1w1piecephi4}
\eea
after series expanding in $\epsilon$ (the integrals converge for $\epsilon > 2$ but we can analytically continue the result to smaller $\epsilon$). As in the case of $\lambda \phi^3$ in 6D, the $\epsilon \rightarrow 0$ divergence can be canceled with the following choice of counterterms,
\bea
    \delta_m \ = \ 0 \, , \quad \frac{1}{1+\delta_r} \ = \ {\rm Exp} \[ \frac{\lambda^2}{6144 \pi^4}\left(\frac{2}{\epsilon}+\gamma+\ln(4\pi)\right) \] ,
\label{eq:d4ctww}
\eea
where, as before, we have used the $\overline{\rm MS}$ renormalization scheme. At second order in $\lambda$, we find that $\delta_r = -\frac{\lambda^2}{6144 \pi^4} \[ \frac{2}{\epsilon} + \gamma + \ln(4\pi) \]$, and thus the counterterms here are similar (but not identical) to those found using in-out in eq.\ (\ref{eq:phi4ct}). Substituting the remaining terms from eq.\ (\ref{eq:c1w1piecephi4}) into eqs.\ (\ref{eq:ww2p3}) and (\ref{eq:cksolww}) then gives us the final connected two-point correlation at two-loops,
\bea
    \big\la \hPhi_{\vk}(t) \hPhi_{\vk'}(t') \big\ra_c & = & (2\pi)^3 \delta^3 \big( \vk+\vk' \big) \frac{e^{-i |\vk| (t-t')}}{2|\vk|} \, {\rm Exp} \Bigg[ -\frac{\lambda ^2}{3072 \pi^4} \nn \\
    & & \qquad \times \, \( \ln \left\{ \mu  (t-t') \right\} + 1 + \frac{i}{2|\vk|} (t-t')^{-1} + \frac{i\pi}{2} \) \Bigg] \, .
\label{eq:ww2pactphi4}
\eea
The decay here is an exact exponential of the (real) secular divergence that we found in eq.\ (\ref{eq:phi42pft4d}), again with extra terms that are either subdominant in the late-time limit or phase factors. As we found for $\lambda\phi^3$ in 6D in eq.\ (\ref{eq:ww2pact6d}), the correlator no longer diverges as $\mu (t-t') \rightarrow \infty$ and instead shows a polynomial decay, again suggesting a successful late-time resummation of the perturbative result.


\section{Geometric series resummation in in-out vs. the WW method}
\label{sec:GRWW}

In this section, we briefly discuss how the geometric series resummation that can be performed for in-out correlators, and can be found in standard textbooks, compares to results obtained using the WW resummation method.
As noted below eq.\ (\ref{eq:2pft4d}), we did not use in-out to its fullest extent, since we did not perform a geometric series resummation before calculating the inverse Fourier transform to the time domain. The reason for this was that the inverse Fourier transform turned out to not be doable. It is doable, however, for the ${\cal O}(\lambda)$ one-loop correction in a massive $\lambda\phi^4$ theory. It is instructive to look at this case as the final result does not diverge in the late-time limit but rather agrees with the corresponding resummation obtained using the WW method (also at first order in $\lambda$), suggesting that the WW method performs a similar resummation away from equilibrium. We will thus consider a $\lambda\phi^4$ theory in 4D, with the Lagrangian density of eq. (\ref{eq:lag4}) and with $m \ne 0$, and focus on the one-loop contribution to the two-point correlator shown in fig.\ \ref{fig:2p1l}. As shown in eq.\ (\ref{eq:phi41loop}), the one-loop contribution to the Green's function is given by
\bea
    G_{F,{\rm 1-loop}}(k) & = & -\frac{i\lambda}{2} \int_p G_F(k) G_F(p) G_F(k) - i \( \delta_m + k^2 \delta_r \) G_F(k) G_F(k) \, .
\label{eq:phi41loop2}
\eea
The loop integral $\frac{\lambda}{2} \int_p G_F(p)$ was calculated in appendix\ \ref{app:31l}, and using the result in eq.\ (\ref{eq:phi3ct1p}) for $d = 4+\epsilon$, we find that
\bea
    G_{F,{\rm 1-loop}}(k) & = & \frac{i\lambda m^2}{32\pi^2} \[ 1 + \ln \( \frac{\mu^2}{m^2} \) \] G_F^2(k) \, ,
\eea
with the following choice of $\overline{\text{MS}}$ counterterms,
\bea
    \delta_m \ = \ -\frac{\lambda m^2}{32\pi^2} \[ \frac{2}{\epsilon} + \gamma - \ln(4\pi) \] , \quad \delta_r \ = \ 0 \, .
\eea
We next perform a resummation of the geometric series in all 1PI contributions to the propagator, which gives
\bea
    G_{F,{\rm resum}}(k) & = & \frac{i}{-k^2 - m^2 + \frac{\lambda m^2}{32\pi^2} \[ 1 + \ln\( \frac{\mu^2}{m^2} \) \] + i\epsilon} \, .
\eea
The positive $\epsilon$ here is different from the one in the counterterm and picks out the correct poles of the propagator. We now plug this expression into eq.\ (\ref{eq:ftfinal}) to perform the inverse Fourier transform over $k_0$, again choosing $t>t'$. The full resummed connected two-point correlation then becomes
\bea
    \big\la \hPhi_{\vk}(t) \hPhi_{\vk'}(t') \big\ra_{c,\rm{resum}} & = & (2\pi)^3 \delta^3 \big( \vk+\vk' \big) \frac{e^{-i \omega_k \sqrt{1 - \frac{\lambda m^2}{32\pi^2 \omega_k} \[ 1 + \ln\( \frac{\mu^2}{m^2} \) \]}(t-t')}}{2\omega_k \sqrt{1 - \frac{\lambda m^2}{32\pi^2 \omega_k} \[ 1 + \ln\( \frac{\mu^2}{m^2} \) \]}} \, ,
\label{eq:ftresum}
\eea
where $\omega_k = \big( |\vk|^2 + m^2 \big)^{1/2}$. The above expression is well-defined in the late-time limit (while its Taylor expansion diverges in this limit) and we thus expect it to be the correct resummation of the two-point correlator at this order. We will next compare this with the corresponding result obtained using the WW method.

With the WW method, the connected two-point correlator for a $\lambda\phi^4$ theory is given by eq.\ (\ref{eq:ww2p3}) with $C_{\vk'}(t,t')$ given by eq.\ (\ref{eq:cksolww}), as noted before. In fact we only need to calculate the first term in eq.\ (\ref{eq:cksolww}) since the second term contributes at second order. This is given by
\bea
    -\frac{i}{V} \int_{t'}^t \d t_1 \, \big\la 1_{\vk'} \big| \hat{H}_{i,I}(t_1) \big| 1_{\vk'} \big\ra & = & \frac{-i\lambda\mu^{4-d}}{2} \int_{t'}^{t} \d t_1 \int_{\vp} f^>_{k'}(t_1) f^<_{k'}(t_1) f^>_{p}(t_1) f^<_{p}(t_1) \nn \\
    & = & \frac{-i\lambda \mu^{4-d} (t-t')}{2^{d+1} \pi^{(d-1)/2} \Gamma \[ (d-1)/2 \] \omega_{k}} \int_0^{\infty} \d |\vp| \frac{|\vp|^{d-2}}{\omega_p} \nn \\
    & = & \frac{-i\lambda m^2 (t-t')}{2^{d+2} \pi^{d/2} \omega_{k}} \( \frac{\mu}{m} \)^{4-d} \Gamma \( 1-\frac{d}{2} \) .
\label{eq:c1w1piecephi41loop}
\eea
We can regularize the integral by setting $d=4+\epsilon$ and use eqs.\ (\ref{eq:wwctdeltam}) and (\ref{eq:wwctdeltar}) to find the appropriate counterterms that cancel the $\epsilon \rightarrow 0$ divergence,
\bea
    \delta_m \ = \ -\frac{\lambda m^2}{32\pi^2} \[ \frac{2}{\epsilon} + \gamma - \ln(4\pi) \] , \quad \delta_r \ = \ 0 \, .
\label{eq:d4ctww}
\eea
Substituting the remaining piece from eq.\ (\ref{eq:c1w1piecephi41loop}) into eqs.\ (\ref{eq:ww2p3}) and (\ref{eq:cksolww}) then gives us the final connected two-point correlation at one-loop,
\bea
    \big\la \hPhi_{\vk}(t) \hPhi_{\vk'}(t') \big\ra_{c,{\rm WW}} & = & (2\pi)^3 \delta^3 \big( \vk+\vk' \big) \frac{e^{-i\omega_{k}(t-t')}}{2\omega_{k}}e^{i\frac{\lambda m^2}{64\pi^2\omega_{k}} \left[1+\ln\left(\frac{\mu^2}{m^2}\right)\right](t-t')} \, . 
\eea
On comparing this expression to that in eq.\ (\ref{eq:ftresum}), we see that the WW method captures the same late-time behavior as the resummation of all 1PI diagrams to one-loop order does in an in-out calculation, after expanding around small $\lambda$ to first order.

\begin{figure}[!t]
\begin{center}
	\includegraphics[scale=0.6]{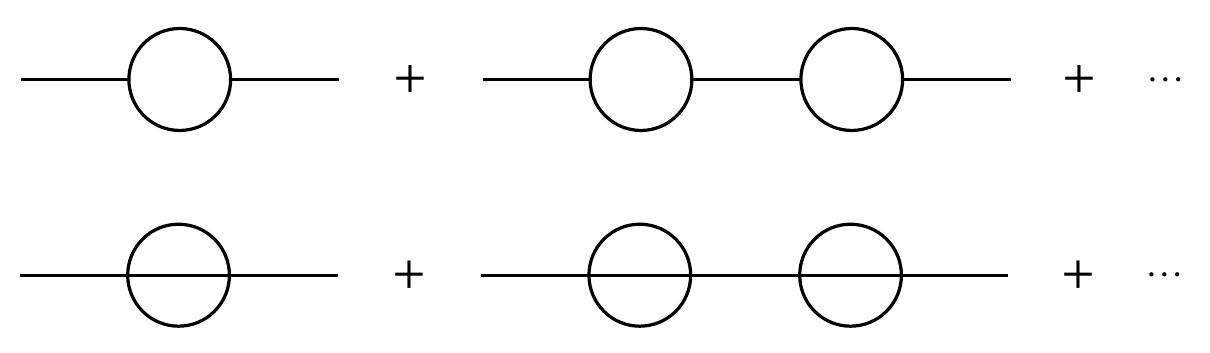}
	\caption{(Top) The geometric series of ${\cal O}(\lambda^2)$ 1PI diagrams that we {\it expect} is being resummed by the WW method when calculating the two-point correlation in a $\lambda\phi^3$ theory. (Bottom) A similar geometric series of (non-vanishing) diagrams in a $\lambda\phi^4$ theory.}
\label{fig:RGWW}
\end{center}
\end{figure}

On the basis of the above example and our results in sections\ \ref{sec:inout} and \ref{sec:ww}, we {\it suspect} that the WW resummation method captures the late-time behavior that the geometric resummation of 1PI diagrams in in-out gives at the same order in perturbation theory. For the massless $\lambda\phi^3$ and $\lambda\phi^4$ theories considered in this paper, we thus believe that the WW method resums the diagrams shown in fig.\ \ref{fig:RGWW}. Exactly how this works mathematically is unclear to us, but we expect that the poles of the propagator are {\it shifted} in the case of $\lambda\phi^3$ in 4D and {\it rescaled} in the cases of $\lambda\phi^3$ in 6D and $\lambda\phi^4$ in 4D, such that the inverse Fourier transform yields an exponential decay for the former and a polynomial decay for the latter. It is important to note, however, that the WW method is still an out-of-equilibrium generalization of the geometric resummation of in-out correlators and so the two results are expected to match only in cases where one can perform a geometric resummation, for example, in Minkowski spacetime and initializing the field in the vacuum of the free theory. We also note that the truncated Hilbert space in the WW method is what restricts the type of diagrams that it resums and one would need to keep more states in the Hilbert space to consistently go to higher orders in perturbation theory.

\section{Discussion}
\label{sec:disc}

Loop corrections in time-dependent interacting quantum field theories are often fraught with late-time divergences. While the divergences may indicate real physical effects in the case of time-dependent spacetimes, in Minkowski spacetime they should simply be an artifact of perturbation theory. In this paper, we were interested in the late-time divergences that arise from loop corrections to the unequal-time two-point correlator in different massless self-interacting scalar quantum field theories on a Minkowski background. We found late-time perturbative results using the standard techniques of in-out perturbation theory and performed late-time resummations using the WW resummation method.

We first considered a $\lambda\phi^3$ interaction in 4D, where $\lambda$ has the dimensions of mass, and found that the perturbative result grows linearly in time. The WW-based result, on the other hand, decays exponentially in time, which is likely an artifact of there being a correlation length proportional to $\lambda^{-1}$ in the system. We next considered two interactions where $\lambda$ is dimensionless, namely a $\lambda\phi^3$ interaction in 6D and a $\lambda\phi^4$ interaction in 4D. In both cases, we found that the perturbative result grows logarithmically in time. The WW-based result, on the other hand, decays polynomially in time, which, in the absence of any scale in the problem to set a correlation length, is similar to the behavior of a system at a critical point \cite{Shankar:1993pf,Rai:2021vvq}. Our results are thus suggestive of `universal' late-time behavior that depends on the dimensions of the interaction strength. We lastly highlighted the type of diagrams that the WW method resums by comparing to the geometric series resummation of in-out correlators, though a more in-depth understanding of this is left to future work.

We note again that we set the mass parameter in the Lagrangian to zero, which does not necessarily imply that the physical mass vanishes. In the cases of $\lambda \phi^3$ in 6D and $\lambda \phi^4$ in 4D, this is indeed correct since the physical mass is proportional to the Lagrangian mass parameter. In the case of $\lambda \phi^3$ in 4D, however, the physical mass is non-zero at second order in $\lambda$, even with the Lagrangian mass parameter being set to zero. We also note that we left our final results for the two-point correlator in terms of an arbitrary renormalization parameter $\mu$. The standard procedure would be to allow the Lagrangian parameters to also depend on $\mu$, define an anomalous dimension and beta function, and use them to express all observables in terms of real physical parameters. We were only interested in the late-time behavior of correlation functions in this paper, however, and have thus forgone this procedure.

\acknowledgments

We especially thank Raquel Ribeiro for initial collaboration and Daniel Boyanovsky for many insightful discussions and comments on an earlier version of this paper. We also thank Yi-Zen Chu, John Collins, Liam Fitzpatrick, Archana Kamal, Sohyun Park, and Sarah Shandera for useful conversations. N.~A. and S.~C. were supported by the Department of Energy under award DE-SC0019515. S.~B. was supported by the National Science Foundation under award PHY-1505411, the Eberly research funds of Penn State, and the Urania E. Stott Fund of The Pittsburgh Foundation.


\appendix
\section{\texorpdfstring{$\lambda\phi^3$}{lp3app} in 4D and 6D: Loop integrals in eqs.\ (\ref{eq:linct}) and (\ref{eq:phi3loop})}
\label{app:31l}

\renewcommand{\theequation}{A\arabic{equation}}
\setcounter{equation}{0}

In this appendix, we calculate the one-loop integrals that appear when calculating the one-point and two-point functions in a $\lambda\phi^3$ interacting theory in 4D and 6D, given in eqs.\ (\ref{eq:linct}) and (\ref{eq:phi3loop}), respectively. These calculations can be found in standard textbooks and are reproduced here simply to make it easy to refer to their results in the main text.

\subsubsection*{(i) Loop integral in eq.\ (\ref{eq:linct})}

We first consider the integral
\bea
    Y & = & \frac{\lambda}{2} \int_p G_F(p) \, ,
\eea
and evaluate it in $d$ dimensions, setting $d = 4+\epsilon$ or $d = 6+\epsilon$ at the end of the calculation, where $\epsilon$ is a dimensional regulator. We also replace $\lambda$ with $\lambda \mu^{4-d}$ or $\lambda \mu^{6-d}$, respectively, $\mu$ being a parameter with dimensions of mass, to keep its dimensions constant; we replace it with $\lambda \mu^{-\epsilon}$ below so that the calculation covers both cases. Now substituting the Feynman Green's function from eq.\ (\ref{eq:GF}), with a non-zero mass, and dropping the $\epsilon$ in the Green's function (keeping the positions of the poles in mind, so as to not cross them when doing a Wick rotation) gives
\bea
    Y & = & -\frac{i\lambda\mu^{-\epsilon}}{2} \int \frac{\d^d p}{(2\pi)^d} \, \frac{1}{p^2 + m^2} \, .
\eea
We next perform a Wick rotation on the contour counterclockwise by $90^0$, and define a Euclidean time coordinate as $p^0 = i \bar{p}_d$, with $p_i = \bar{p}_i$ for $i = 1, \ldots, d-1$. The angular part of the resulting $\bar{p}$ integral gives $\int \d\Omega_d = \frac{2\pi^{d/2}}{\Gamma(d/2)}$ and we are left with
\bea
    Y & = & \frac{\lambda\mu^{-\epsilon}}{2^d \pi^{d/2} \Gamma(d/2)} \int_0^{\infty} \d p \, \frac{p^{d-1}}{p^2 + m^2} \nn \\
    & = & \frac{\lambda\mu^{-\epsilon} m^{d-2}}{2^{d+1} \pi^{d/2-1}} \frac{\csc(d\pi/2)}{\Gamma(d/2)}
\eea
for $d < 2$. We can analytically continue this result to higher $d$ since we know that the interaction is renormalizable in both 4D and 6D. We see already that $Y$ vanishes in the massless limit, but let us write it out explicitly by plugging in specific values for $d$ and series expanding in $\epsilon$,
\bea
    Y & = &
	\begin{cases}
        - \frac{\lambda m^2}{32 \pi^2} \[ 1 - \gamma + \ln \( \frac{4\pi\mu^2}{m^2} \) \] + \frac{\lambda m^2}{16 \pi^2 \epsilon} & (d = 4 + \epsilon) \, ,
\vspace{0.25cm} \\
        \frac{\lambda m^4}{256 \pi^3} \[ \frac{3}{2} - \gamma + \log \( \frac{4\pi\mu^2}{m^2} \) \] - \frac{\lambda m^4}{128 \pi^3 \epsilon} & (d = 6 + \epsilon) \, ,
	\end{cases}
\label{eq:phi3ct1p}
\eea
where we have separated out the leading-order divergence in $\epsilon$. As mentioned above, $Y$ vanishes in the massless limit that we consider in this paper.

\subsubsection*{(ii) Loop integral in eq.\ (\ref{eq:phi3loop})}

We next consider the integral
\bea
    I_{\lambda\phi^3}(k) & = & \frac{\lambda^2}{2} \int_p G_F(p) G_F(k-p) \, ,
\eea
and again evaluate it in $d$ dimensions, setting $d = 4+\epsilon$ or $d = 6+\epsilon$ at the end of the calculation. We also replace $\lambda$ with $\lambda \mu^{2-d/2}$ or $\lambda \mu^{3-d/2}$, respectively, to keep its dimensions constant; we replace it with $\lambda \mu^{-\epsilon/2}$ below so that the calculation covers both cases. Now substituting the Feynman Green's function from eq.\ (\ref{eq:GF}), setting the mass to zero, and dropping the $\epsilon$s in the Green's functions gives
\bea
    I_{\lambda\phi^3}(k) & = & -\frac{\lambda^2 \mu^{-\epsilon}}{2} \int \frac{\d^dp}{(2\pi)^d} \, \frac{1}{p^2(k-p)^2} \, .
\eea
We can write the denominator in the integrand using Feynman's parametrization as
\bea
    \frac{1}{p^2(k-p)^2} & = & \int_0^1 \d x \, \frac{1}{\[ (1-x)p^2 + x(k-p)^2\]^2} \nn \\
    & = & \int_0^1 \d x \, \frac{1}{\[ q^2 + x(1-x)k^2 \]^2} \, ,
\label{eq:feynpar}
\eea
where $q = p - xk$. We can now replace the $p$ integral with an integral over $q$, perform a Wick rotation as before on the contour counterclockwise by $90^0$, and define a Euclidean time coordinate as $q^0 = i \bar{q}_d$, with $q_i = \bar{q}_i$ for $i = 1, \ldots, d-1$. The angular part of the resulting $\bar{q}$ integral gives $\int \d\Omega_d = \frac{2\pi^{d/2}}{\Gamma(d/2)}$ and we are left with
\bea
    I_{\lambda\phi^3}(k) & = & -\frac{i\lambda^2 \mu^{-\epsilon}}{2^d \pi^{d/2} \Gamma(d/2)} \int_0^1 \d x \int_0^{\infty} \d \bar{q} \, \frac{\bar{q}^{d-1}}{\[ \bar{q}^2 + x(1-x)k^2 \]^2} \nn \\
    & = & \frac{i\lambda^2 \mu^{-\epsilon} (d-2)k^{d-4}}{2^{d+2} \pi^{d/2-1}} \frac{\csc(d\pi/2)}{\Gamma(d/2)} \int_0^1 \d x \, [x(1-x)]^{d/2-2}
\eea
for $d<4$, which we will again analytically continue to higher $d$. Lastly, performing the $x$ integral gives
\bea
    I_{\lambda\phi^3}(k) & = & \frac{i\lambda^2 \mu^{-\epsilon} (d-2) k^{d-4}}{2^{2d-1} \pi^{(d-3)/2}} \frac{\csc(d\pi/2) \Gamma(d/2-1)}{\Gamma(d/2) \Gamma[(d-1)/2]} \, .
\label{eq:phi3loopbeforeexp}
\eea
We can now set specific values for $d$ and series expand in $\epsilon$ to obtain the following expressions in the two cases of interest,
\bea
	I_{\lambda\phi^3}(k) & = &
	\begin{cases}
		\frac{i\lambda^2}{16\pi^2 \epsilon} - \frac{i\lambda^2}{32\pi^2} \[ 2 - \gamma + \log \( \frac{4\pi\mu^2}{k^2} \) \] & (d = 4 + \epsilon) \, ,
\vspace{0.25cm} \\
		-\frac{i\lambda^2 k^2}{384\pi^3 \epsilon} + \frac{i\lambda^2 k^2}{768\pi^3} \[ \frac{8}{3}-\gamma + \log \( \frac{4\pi\mu^2}{k^2} \) \] & (d = 6 + \epsilon) \, ,
	\end{cases}
\label{eq:phi3loop4d6d}
\eea
where we have separated out the leading-order divergence in $\epsilon$.


\section{\texorpdfstring{$\lambda\phi^4$}{lp4app} in 4D: Loop integral in eq.\ (\ref{eq:sunsetloop})}
\label{app:42l}

\renewcommand{\theequation}{B\arabic{equation}}
\setcounter{equation}{0}

In this appendix, we calculate the two-loop integral that appears when calculating the two-point function in a $\lambda\phi^4$ interacting theory in 4D, given in eq.\ (\ref{eq:sunsetloop}). This calculation can also be found in standard textbooks and is reproduced here simply to make it easy to refer to its result in the main text.

We consider the integral
\bea
    I_{\lambda\phi^4}(k) & = & \frac{\lambda^2}{6} \int_p\int_{p'} G_F(p)G_F(p')G_F(k-p-p') \, ,
\eea
and evaluate it in $d$ dimensions, setting $d = 4+\epsilon$ at the end of the calculation. We also replace $\lambda$ with $\lambda \mu^{4-d}$ to keep it dimensionless. Now substituting the Feynman Green's function from eq.\ (\ref{eq:GF}), setting the mass to zero, and dropping the $\epsilon$s in the Green's functions gives
\bea
    I_{\lambda\phi^4}(k) & = & \frac{i\lambda^2\mu^{8-2d}}{6} \int \frac{\d^dp}{(2\pi)^d} \frac{\d^dp'}{(2\pi)^d} \, \frac{1}{p^2p'^2(k-p-p')^2} \, .
\eea
We can write the denominator in the integrand using Feynman's parametrization as in appendix\ \ref{app:31l},
\bea
    \frac{1}{p'^2(k-p-p')^2} & = & \int_0^1 \d x \, \frac{1}{\[ (1-x)p^{\prime 2} + x(k-p-p')^2\]^2} \nn \\
    & = & \int_0^1 \d x \, \frac{1}{\[ q'^2 + x(1-x)(k-p)^2 \]^2} \, ,
\label{eq:feynpar}
\eea
where $q' = p' - x(k-p)$. We can now replace the $p'$ integral with an integral over $q'$, perform a Wick rotation, and convert to a Euclidean integral over $\bar{q}'$ as in appendix\ \ref{app:31l}. Also performing the angular part of the $\bar{q}'$ integral gives
\bea
    I_{\lambda\phi^4}(k) & = & -\frac{\lambda^2\mu^{8-2d}}{3 \cdot 2^d \pi^{d/2} \Gamma(d/2)} \int \frac{\d^dp}{(2\pi)^d} \int_0^1 \d x \int_0^{\infty} \d \bar{q}' \, \frac{\bar{q}'^{d-1}}{p^2 \[ \bar{q}'^2 + x(1-x)(k-p)^2 \]^2} \nn \\
    & = & \frac{\lambda^2\mu^{8-2d} (d-2)}{3 \cdot 2^{d+2} \pi^{d/2-1}} \frac{\csc(d\pi/2)}{\Gamma(d/2)} \int \frac{\d^dp}{(2\pi)^d} \int_0^1 \d x \, \frac{\[x(1-x)\]^{d/2-2}}{p^2 (k-p)^{4-d}}
\label{eq:Ilphi4p}
\eea
for $d < 4$. We can analytically continue this result to higher $d$ since we know that the interaction is renormalizable in 4D. We can also use Feynman's parametrization again to write the denominator in eq.\ (\ref{eq:Ilphi4p}) as
\bea
    \frac{1}{p^2 (k-p)^{4-d}} & = & \frac{\Gamma(3-d/2)}{\Gamma(2-d/2)} \int_0^1 \d y \, \frac{y^{1-d/2}}{\[ q^2 + y(1-y) k^2 \]^{3-d/2}} \, ,
\eea
where $q = p - yk$. As before, we can replace the $p$ integral with an integral over $q$, perform a Wick rotation, and convert to a Euclidean integral over $\bar{q}$. After performing the $\bar{q}$ angular integral, we find that
\bea
    I_{\lambda\phi^4}(k) & = & \frac{i\lambda^2\mu^{8-2d} (d-2)}{3 \cdot 2^{2d+1} \pi^{d-1}} \frac{\csc(d\pi/2) \Gamma(3-d/2)}{\Gamma^2(d/2) \Gamma(2-d/2)} \nn \\
    & & \qquad \times \, \int_0^1 \d x \int_0^1 \d y \int_0^{\infty} \d \bar{q} \, \frac{\[x(1-x)\]^{d/2-2} y^{1-d/2} \bar{q}^{d-1}}{\[ q^2 + y(1-y) k^2 \]^{3-d/2}} \nn \\
    & = & \frac{i\lambda^2\mu^{8-2d} (d-2) k^{2d-6}}{3 \cdot 2^{2d+2} \pi^{d-1}} \frac{\csc(d\pi/2) \Gamma(3-d)}{\Gamma(d/2) \Gamma(2-d/2)} \int_0^1 \d x \int_0^1 \d y \, \[xy(1-x)\]^{d/2-2} (1-y)^{d-3} \nn \\
\eea
for $d < 3$, which we again analytically continue to higher $d$. Lastly, performing the $x$ and $y$ integrals gives
\bea
    I_{\lambda\phi^4}(k) & = & \frac{i\lambda^2\mu^{8-2d} (d-2) k^{2d-6}}{3 \cdot 2^{2d+2} \pi^{d-1}} \frac{\csc(d\pi/2) \Gamma(3-d) \Gamma^3(d/2-1)}{\Gamma(d/2) \Gamma(2-d/2) \Gamma(3d/2-3)} \, .
\eea
We can now set $d = 4+\epsilon$ and series expand in $\epsilon$ to obtain
\bea
    I_{\lambda\phi^4}(k) & = & -\frac{i \lambda^2 k^2}{3072 \pi ^4 \epsilon} + \frac{i \lambda^2 k^2}{3072 \pi^4} \[ \frac{13}{4} - \gamma + \log \( \frac{4\pi\mu^2}{k^2} \) \] ,
\label{eq:phi4loop4d}
\eea
where we have separated out the leading-order divergence in $\epsilon$.


\section{\texorpdfstring{$\lambda\phi^3$}{lp3app} in 4D and 6D: Vacuum state amplitude in eq.\ (\ref{eq:ww2pc0})}
\label{app:C0}

\renewcommand{\theequation}{C\arabic{equation}}
\setcounter{equation}{0}

In this appendix, we calculate the vacuum state amplitude $C_0(t)$ that appears when calculating the two-point function in a $\lambda\phi^3$ interacting theory in 4D and 6D using the WW resummation method, in eq.\ (\ref{eq:ww2pc0}). Using eq.\ (\ref{eq:casolft}), $C_0(t)$ is given by
\bea
    C_0(t) & = & {\rm Exp} \[ -i \int_{t_0}^t \d t_1 \, \big\la 0 \big| \hat{H}_{i,I}(t_1) \big| 0 \big\ra - \int_{t_0}^t \d t_1 \int_{t_0}^{t_1} \d t_2 \, \Sigma_0(t_1,t_2) \] ,
\label{eq:c0def}
\eea
and we are specifically interested in the interaction Hamiltonian $\lambda {\,:}\hat{\Phi}^3{:\,}$. The first term in the exponential vanishes for this interaction and we thus only need to calculate
\bea
    \Sigma_{0}(t_1,t_2) & = & \frac{1}{6} \int_{\vp_1} \int_{\vp_2}\int_{\vp_3} \big\la 0 \big| \hat{H}_{i,I}(t_1) \big| 1_{\vp_1}, 1_{\vp_2},1_{\vp_3} \big\ra \big\la 1_{\vp_1}, 1_{\vp_2}, 1_{\vp_3} \big| \hat{H}_{i,I}(t_2) \big| 0 \big\ra \nn\\
    & = & \frac{\lambda^2 V}{6} \int_{\vp_1} \int_{\vp_2} \frac{e^{-i|\vp_1|(t_1-t_2)}}{2|\vp_1|} \frac{e^{-i|\vp_2|(t_1-t_2)}}{2|\vp_2|} \frac{e^{-i|\vp_1+\vp_2|(t_1-t_2)}}{2|\vp_1+\vp_2|} \, , \quad
\label{sigma0c} \, 
\eea
where the volume factor $V$ is expected to appear due to the extensivity of the vacuum energy \cite{Boyanovsky:2011xn}. We now change the time integrals to $\int \d \tau \int \d \Delta$, with $\tau = t_1 - t_0$ and $\Delta = t_1 - t_2$, and rewrite eq.\ (\ref{sigma0c}) in terms of the new time variables as
\bea
    \Sigma_{0}(\tau,\Delta) & = & \frac{\lambda^2 V}{(3) 2^{2d} \pi^{d-1/2} \Gamma[(d-1)/2] \Gamma(d/2-1)} \int_0^{\infty} \d|\vp_1|\ |\vp_1|^{d-4}\int_0^{\infty} \d|\vp_2|\ |\vp_2|^{d-4} \nn \\
    & & \quad \times \int_{||\vp_1|-|\vp_2||}^{|\vp_1| + |\vec{p}_2|} \d |\vq| \, \[ 1 - \frac{\( |\vp_1|^2 + |\vp_2|^2 - |\vq|^2 \)^2}{4 |\vp_1|^2 |\vp_2|^2} \]^{d/2-2} e^{-i(|\vp_1|+|\vp_2|+|\vq|) \Delta} \, ,
\label{sigma0cp2}
\eea
where we have used $\int \d^{d} x = \frac{2\pi^{(d-1)/2}}{\Gamma [(d-1)/2]} \int_0^{\infty} \d x \int_{-1}^1 \d \cos\theta \, x^{d-1} \sin^{d-3}\theta$, defined $\vq = \vp_1+\vp_2$, so that $|\vq| = \( |\vp_1|^2 + |\vp_2|^2 + 2|\vp_1| |\vp_2| \cos\theta_2 \)^{1/2}$, and changed the integral over $\cos\theta_2$ to $|\vq|$ using $\d \cos\theta_2 = \frac{|\vq|}{|\vp_1| |\vp_2|} \, \d |\vq|$.

We will now specialize to either $d=4$ or $d=6$ to obtain $\int_0^{t-t_0} \d \tau \int_0^{\tau} \d \Delta \, \Sigma_{0}(\tau,\Delta)$ and thus $C_{0}(t)$. To regulate the loop integral in eq.\ (\ref{sigma0cp2}), we convert $\Delta$ into a Euclidean time coordinate by defining $\Delta = -i\bar{\Delta}$ and use dimensional regularization on $\bar{\Delta}$. In other words, we replace $\int \d \bar{\Delta}$ with $\mu^{\epsilon} \int \d^{1+\epsilon} \bar{\Delta} = \mu^{\epsilon} \frac{\pi^{(1+\epsilon)/2}}{\Gamma[(1+\epsilon)/2]} \int \d \bar{\Delta} \, \bar{\Delta}^{\epsilon}$ and set $d=4$ or $d=6$ in eq.\ (\ref{sigma0cp2}). For $d=4$, the double time integral of $\Sigma_{0}(\tau,\Delta)$ then becomes
\bea
    \int_0^{t-t_0} \d\tau \int_0^{\tau} \d \Delta \, \Sigma_0(\tau,\Delta) & = & \frac{-i\lambda^2 \mu^{\epsilon} V}{384\pi^4} \frac{\pi^{(1+\epsilon)/2}}{\Gamma[(1+\epsilon)/2]} \int_{0}^{t-t_0} \d\tau \int_0^{i\tau} \d\bar{\Delta} \, \bar{\Delta}^{\epsilon} \nn \\
    & & \qquad \times \int_0^{\infty} \d|\vp_1| \int_0^{\infty} \d|\vp_2| \int_{||\vp_1| - |\vp_2||}^{|\vp_1| + |\vp_2|} \d |\vq| \,  e^{-(|\vp_1|+|\vp_2|+|\vq|) \bar{\Delta}} \nn \\
    & = & \frac{i\lambda^2 V}{3072\pi^4 (t-t_0)} + {\cal O}(\epsilon) \, .
\label{4dphase}
\eea
We thus obtain $C_0(t) = {\rm Exp} \[ -\frac{i\lambda^2 V}{3072\pi^4 (t-t_0)} \]$, which is simply a phase factor and in fact goes to unity in the late-time limit. Similarly, on setting $d=6$ instead in eq.\ (\ref{sigma0cp2}) and taking its double time integral, we obtain $C_0(t) = {\rm Exp} \[ -\frac{i\lambda^2 V}{737280\pi^6 (t-t_0)^5} \]$, which is also simply a phase factor and goes to unity in the late-time limit.

\section{\texorpdfstring{$\lambda\phi^4$}{lp3app} in 4D: Vacuum state amplitude in eq.\ (\ref{eq:ww2pc0})}
\label{app:D0}

\renewcommand{\theequation}{D\arabic{equation}}
\setcounter{equation}{0}

In this appendix, we calculate the vacuum state amplitude $C_0(t)$ that appears when calculating the two-point function in a $\lambda\phi^4$ interacting theory in 4D using the WW resummation method, in eq.\ (\ref{eq:ww2pc0}). Similar to the $\lambda\phi^3$ case, $C_0(t)$ is given by
\bea
    C_0(t) & = & {\rm Exp} \[ -i \int_{t_0}^t \d t_1 \, \big\la 0 \big| \hat{H}_{i,I}(t_1) \big| 0 \big\ra - \int_{t_0}^t \d t_1 \int_{t_0}^{t_1} \d t_2 \, \Sigma_0(t_1,t_2) \] ,
\label{eq:c0def4}
\eea
and we are now specifically interested in the interaction Hamiltonian $\lambda \hat{\Phi}^4$. On using eq.\ (\ref{eq:cq}) for $\hPhi_{\vk,I}(t)$, the first term in the exponential becomes
\bea
    -i \int_{t_0}^t \d t_1 \, \big\la 0 \big| \hat{H}_{i,I}(t_1) \big| 0 \big\ra & = & -\frac{iV}{8}\int_{t_0}^t \d t_1\int_{\vp_1}\int_{\vp_2} \frac{1}{2|\vp_1|}\frac{1}{2|\vp_2|} \, ,
\eea
where we again find a volume factor, as is typical for vacuum corrections. Each of the two loop integrals is identical to that encountered in eqs.\ (\ref{eq:linct}) and (\ref{eq:c1w1piecephi41loop}) and, as found there, vanishes in the massless limit. We thus only need to calculate
\bea
    \Sigma_{0}(t_1,t_2) & = & \frac{1}{24} \int_{\vp_1} \int_{\vp_2}\int_{\vp_3}\int_{\vp_4} \big\la 0 \big| \hat{H}_{i,I}(t_1) \big| 1_{\vp_1}, 1_{\vp_2},1_{\vp_3},1_{\vp_4} \big\ra \big\la 1_{\vp_1}, 1_{\vp_2}, 1_{\vp_3} , 1_{\vp_4} \big| \hat{H}_{i,I}(t_2) \big| 0 \big\ra \nn\\
    & = & \frac{\lambda^2 V}{24} \int_{\vp_1} \int_{\vp_2}\int_{\vp_3}  \frac{e^{-i|\vp_1|(t_1-t_2)}}{2|\vp_1|} \frac{e^{-i|\vp_2|(t_1-t_2)}}{2|\vp_2|}\frac{e^{-i|\vp_3|(t_1-t_2)}}{2|\vp_3|} \frac{e^{-i|\vp_1+\vp_2+\vp_3|(t_1-t_2)}}{2|\vp_1+\vp_2+\vp_3|} \, . \quad
\label{sigma0c4} \, 
\eea
We now change the time integrals to $\int \d \tau \int \d \Delta$, with $\tau = t_1 - t_0$ and $\Delta = t_1 - t_2$, and rewrite eq.\ (\ref{sigma0c4}) in terms of the new time variables as
\bea
    \Sigma_{0}(\tau,\Delta) & = & \frac{\lambda^2 V}{(3) 2^{12} \pi^{6}} \int_0^{\infty} \d|\vp_1|\ \int_0^{\infty} \d|\vp_2|\int_{||\vp_1|-|\vp_2||}^{|\vp_1| + |\vp_2|} \d |\vq_2|\int_{0}^{\infty} \d |\vp_3| \nn \\
    & & \quad \times\int_{||\vq_2|-|\vp_3||}^{|\vq_2| + |\vec{p}_3|} \d |\vq_3| \, e^{-i(|\vp_1|+|\vp_2|+|\vp_3|+|\vq_3|) \Delta} \, ,
\label{sigma0cp24}
\eea
where we have set the number of spatial dimensions to three, defined $\vq_2 = \vp_1+\vp_2$ and $\vq_3 = \vq_2+\vp_3$, so that $|\vq_2| = \big( |\vp_1|^2 + |\vp_2|^2 + 2|\vp_1| |\vp_2| \cos\theta_2 \big)^{1/2}$ and $|\vq_3| = \big( |\vq_2|^2 + |\vp_3|^2 + 2|\vq_2| |\vp_3| \cos\theta_3 \big)^{1/2}$, and changed the integrals over $\cos\theta_2$ and $\cos\theta_3$ to $|\vq_2|$ and $|\vq_3|$ using $\d \cos\theta_2 = \frac{|\vq_2|}{|\vp_1| |\vp_2|} \, \d |\vq_2|$ and $\d \cos\theta_3 = \frac{|\vq_3|}{|\vq_2| |\vp_3|} \, \d |\vq_3|$, respectively.

We are interested in $\int_0^{t-t_0} \d \tau \int_0^{\tau} \d \Delta \, \Sigma_{0}(\tau,\Delta)$, which will then give us $C_{0}(t)$. To regulate the loop integral in eq.\ (\ref{sigma0cp24}), we convert $\Delta$ into a Euclidean time coordinate by defining $\Delta = -i\bar{\Delta}$, as before. We then use dimensional regularization on $\bar{\Delta}$, absorbing the fractional dimension fully into the time coordinate, so that the number of spatial dimensions is still three. The double time integral of $\Sigma_{0}(\tau,\Delta)$ then becomes
\bea
    \int_0^{t-t_0} \d\tau \int_0^{\tau} \d \Delta \, \Sigma_0(\tau,\Delta) & = & \frac{-i\lambda^2 \mu^{\epsilon} V}{(3)2^{14}\pi^6} \frac{\pi^{(1+\epsilon)/2}}{\Gamma[(1+\epsilon)/2]} \int_{0}^{t-t_0} \d\tau \int_0^{i\tau} \d\bar{\Delta} \, \bar{\Delta}^{\epsilon-5} \nn \\
    & = & -\frac{i\lambda^2 V}{589824 \pi^6 (t-t_0)^3} + {\cal O}(\epsilon) \, .
\label{4dphase4}
\eea
We thus obtain $C_0(t) = {\rm Exp} \[ \frac{i\lambda^2 V}{589824 \pi^6 (t-t_0)^3} \]$, which is also simply a phase factor and goes to unity in the late-time limit.


\bibliography{references}
\bibliographystyle{JHEP}

\end{document}